\documentclass[12pt]{article}
\usepackage{epsfig,amsfonts,amssymb}
\usepackage{hyperref}
\usepackage{cite}
\input epsf.sty
\usepackage{cite}

\def\ZZZ{{\hbox{ Z\kern-1.6mm Z}}}
\def\RRR{{\hbox{ R\kern-2.4mm R}}}
\def\CCC{{\hbox{ C\kern-2.0mm C}}}
\def\zzz{{\hbox{z\kern-1mm z}}}

\newcommand{\qeq}{{\hbox{=\kern-2.3mm ? \kern.5mm }}}
\renewcommand{\qeq}{=}

\newcommand{\eps}{\epsilon}

\newcommand{\vp}{\varphi}

\newcommand{\VV}{{\cal V}}

\newcommand{\DD}{{\cal D}}

\newcommand{\KK}{{\cal K}}

\newcommand{\HH}{{\cal H}}
\newcommand{\MM}{{\cal M}}
\newcommand{\CC}{{\cal C}}

\newcommand{\OO}{{\cal O}}

\newcommand{\PP}{{\cal P}}

\newcommand{\wt}{\widetilde}
\newcommand{\wh}{\widehat}

\newcommand{\RR}{{\cal R}}

\newcommand{\SSS}{{\cal S}}

\newcommand{\be}{\begin{equation}}
\newcommand{\ee}{\end{equation}}
\newcommand{\ben}{\begin{eqnarray}\displaystyle}
\newcommand{\een}{\end{eqnarray}}

\newcommand{\refb}[1]{(\ref{#1})}
\newcommand{\p}{\partial}
\newcommand{\sectiono}[1]{\section{#1}\setcounter{equation}{0}}

\def\one{{\hbox{ 1\kern-.8mm l}}}
\def\zero{{\hbox{ 0\kern-1.5mm 0}}}

\newcommand{\bea}[1]{\begin{eqnarray}\label{#1} }
\newcommand{\eea}{\end{eqnarray}}

\newcommand{\eqref}{\refb}

\newcommand{\XX}{{\cal X}}

\usepackage{bm}
\usepackage[table]{xcolor}

\begin{document}

\baselineskip 24pt

\begin{center}
{\Large \bf  Gauge Invariant 1PI 
Effective Action for  Superstring Field Theory}

\end{center}

\vskip .6cm
\medskip

\vspace*{4.0ex}

\baselineskip=18pt

\centerline{\large \rm Ashoke Sen}

\vspace*{4.0ex}

\centerline{\large \it Harish-Chandra Research Institute}
\centerline{\large \it  Chhatnag Road, Jhusi,
Allahabad 211019, India}

\vspace*{1.0ex}
\centerline{\small E-mail:  sen@mri.ernet.in}

\vspace*{5.0ex}

\centerline{\bf Abstract} \bigskip

We construct  gauge invariant 1PI effective 
action for the NS sector of type II and heterotic string
field theory. By construction, zero eigenvalues of the kinetic operator of this 
action determine the renormalized physical masses, and 
tree level amplitudes computed from this action 
(after gauge fixing) give the loop corrected
S-matrix elements. Using this
formalism we can give a simple proof of the result 
that the renormalized physical masses
do not
depend on the choice of local coordinate system and locations of picture changing
operators used in defining the off-shell amplitude.
We also eliminate the need for an infrared regulator in dealing with tadpoles of
massless fields.

\vfill \eject

\baselineskip=18pt

\tableofcontents

\sectiono{Introduction}

Construction of a manifestly Lorentz invariant 
field theory for heterotic or type II strings 
has been an open problem. The problem essentially comes from the Ramond (R)
sector -- there is no natural, fully consistent 
candidate for the kinetic term of R sector fields.
In contrast in the Neveu-Schwarz (NS) 
sector one can write down a consistent field theory at the tree
level\cite{wittenssft,9202087,9503099,0109100,0406212,0409018,0911.2962,1201.1761,
1303.2323,1312.1677,1312.2948,1312.7197,1403.0940}. 
However since in quantum theory R sector states propagate
in the loop even if the external states are all NS sector states, absence of a tree
level string field theory including the R sector states 
constitutes a bottleneck in the construction
of a string field theory at the quantum level.

The purpose of this note is to circumvent this problem by introducing the notion of
gauge invariant
one particle irreducible (1PI) effective action for string field theory involving 
NS sector external states. As in conventional field theory,  the full
quantum amplitudes are given by the {\it tree level} amplitudes computed from
this 1PI effective action. Thus if we are working with external states in the NS sector, we 
do not need to worry about the R sector states -- they have already been integrated out
in constructing this 1PI action. 

In a quantum field theory the construction of the 1PI action requires fixing a
gauge, computing the 1PI amplitudes using the propagators and vertices of the
gauge fixed theory, recasting the result in terms of an 1PI
action, and then expressing the result as a sum of a gauge invariant
effective action and a gauge fixing term. It turns out that string theory allows us to
bypass all the steps and directly construct a gauge invariant 1PI action. Of course the
final S-matrix elements computed from this
are still expressed as integrals over the moduli spaces of
Riemann surfaces, and so the construction of this 1PI action does not simplify this
computation. However since the 1PI action can be used to ask questions which
require going off-shell in the intermediate stage, {\it e.g.} the issues of mass
renormalization \cite{1311.1257,1401.7014} 
and vacuum shift \cite{1404.6254} studied recently, 
using the 1PI effective action we can simplify the analysis of these
questions. 

Our construction of the 1PI effective action will be based on the definition of the off-shell
amplitude given in \cite{1408.0571}
using the formalism of picture changing operators 
(PCO's)\cite{FMS,Verlinde:1987sd,lechtenfeld,morozov}. 
On-shell there is also
a more geometric approach based on integration over supermoduli 
space\cite{Belopolsky,9706033,dp,1209.5461,1304.2832,Witten,donagi-witten,1403.5494}.
Off-shell generalizations of this formalism was attempted in \cite{Belopolsky,9706033}, 
but, to our knowledge,
a fully satisfactory formalism does not yet exist. If such a formalism is developed then it
can also in principle be used to give a different construction of the 1PI effective action.

The 1PI action we have introduced here 
leads us to the same definition
of off-shell amplitudes as the ones used in \cite{1311.1257,1401.7014,1404.6254}. 
The main advantage of using
the formalism
of gauge invariant 1PI action is that using this we can simplify the proof that various
physical quantities computed from off-shell amplitudes are independent of the choice
of the local coordinates or the locations of the PCO's used in
defining the off-shell amplitude.  We list below 
some of the concrete results coming out of this
formalism:
\begin{enumerate}
\item In the presence of massless fields, conventional string perturbation theory requires
an infrared cut-off to regulate contributions from separating type degenerations 
at the intermediate stages of calculation\cite{1209.5461}. Our approach based on gauge
invariant 1PI effective action eliminates the need of such an infrared cut-off by
subtracting off the contribution from massless fields to the propagator using a well defined
procedure. 
\item It has been known for a while\cite{catoptric,1209.5461} that if we change the local
coordinate system and/or PCO locations, then in order to keep the physical quantities
(including conventional on-shell S-matrix elements)
unchanged we need to make appropriate shifts of various moduli fields. The formalism
based on 1PI effective action
provides a natural explanation of this result by using the fact that two different 1PI effective
action corresponding to different choices of local coordinate systems and/or PCO
locations are related by a field redefinition and this field redefinition in general produces
a shift in the moduli fields.
\item Ref.\cite{1401.7014} described a general procedure for computing renormalized masses
of physical states using the Siegel gauge kinetic term. However for general states
a direct proof of the fact that
these renormalized masses are invariant under a change of local coordinate system
and/or PCO locations was lacking. We give a proof of this by relating the Siegel gauge
analysis to the analysis of the kinetic operator of the gauge invariant 1PI effective action.
Since under a change in the local coordinates / PCO locations the kinetic
operator of the 1PI action gets conjugated by a non-singular matrix, the locations of its
zero eigenvalues in the momentum space, representing physical states, remain unchanged.
\item Ref.\cite{1404.6254} described a general procedure of computing string theory amplitudes
around a vacuum that is not the perturbative vacuum but is obtained from it by a small
shift of
a scalar field. Using the 1PI effective action one can give a justification of these rules.
Also unlike in the approach described in \cite{1404.6254} where one needs an
infrared regulator at the intermediate stages of the calculation for regulating massless
tadpoles, in the approach based on 1PI effective action we determine the vacuum expectation
values of the massless fields by solving their equations of motion, and hence avoid the
appearance of tadpoles even at the intermediate steps of the calculation.
\end{enumerate}

The rest of the paper is organised as follows. In \S\ref{sback} we review the background
material needed for the construction of the 1PI action. In \S\ref{saction} we describe the
1PI action, its gauge invariance, equations of motions and the 
construction of the vacuum solutions. We also show that the change in the 1PI action 
under a change in the choice of local coordinates at the punctures
and locations of the PCO's used in the construction of the
1PI action can be absorbed into a redefinition of the string fields. Most of the
results in this section are adaptations of the corresponding results in classical bosonic
string field theory\cite{wittensft,saadi,kugo,9206084,aseq,9301097} 
in this new context. In \S\ref{sapp} we apply this formulation to study
mass renormalization and vacuum shift effects in string theory. In particular we 
give a simple argument showing that
the renormalized physical mass and the S-matrix elements in perturbative vacuum  
are
independent of the choice of local coordinate system and the PCO locations used
to construct the 1PI action.
This argument can also be extended to 
the shifted vacuum when the latter is the correct ground state,
Finally in appendix \ref{svert} we extend the analysis of \S\ref{eeffct} to show that even
when we consider two choices of PCO locations which differ in their vertical segment,
there is a field redeinition that relates the corresponding 1PI actions.

\sectiono{Background} \label{sback}

In this section we shall review some of the background material
that goes into the construction
of the 1PI effective action of string field theory.
For definiteness we focus on the NS sector of the heterotic string theory but the
results have straightforward generalization to the NSNS sector of
type II string theories.
We shall follow the conventions of \cite{1408.0571}.
The world sheet theory contains a
matter superconformal field theory with
central charge (26,10), and a ghost system of  total central charge $(-26,-10)$
containing anti-commuting $b$, $c$, $\bar b$, $\bar c$
ghosts and commuting $\beta, \gamma$ ghosts.
The $(\beta,\gamma)$ system can be bosonized as\cite{FMS}
\be \label{eboserule}
\gamma = \eta\, e^{\phi}, \quad \beta= \p\xi \, e^{-\phi}, \quad \delta(\gamma)
= e^{-\phi}, \quad \delta(\beta) = e^\phi\, ,
\ee
where $\xi, \eta$ are  fermions and $\phi$ is a scalar with background charge.
We assign (ghost number, picture number, GSO) quantum numbers to various fields
as follows:
\ben
&& c, \bar c: (1,0,+), \quad b, \bar b: (-1, 0,+), \quad \gamma: (1,0,-), \quad \beta:(-1,0,-),
\nonumber \\
&& \xi: (-1,1,+), \quad \eta: (1,-1,+), \quad
e^{q\phi}: (0, q, (-1)^q)\, .
\een
We do not make any specific choice of the matter superconformal field theory except
that it has a free part describing Minkowski space-time so that we have the notion
of mass spectrum, S-matrix etc. 

We now introduce a vector space $\HH_0$ containing a subset of states in the matter-ghost
conformal field theory satisfying the following conditions:
\be \label{econd}
|s\rangle \in \HH_0 \quad \hbox{iff} \quad b_0^-|s\rangle = 0, \quad L_0^{-}|s\rangle =0\, ,
\quad \eta_0 |s\rangle =0, 
\quad \hbox{picture number of $|s\rangle=-1$}, 
\ee
where
\be
b_0^\pm \equiv(b_0\pm\bar b_0), \quad L_0^\pm\equiv(L_0\pm\bar L_0)\, .
\ee
The off-shell string field is taken to be a state $|\Psi\rangle\in\HH_0$ of ghost number 2.
For later use we also define
\be
c_0^\pm = {1\over 2} (c_0\pm\bar c_0)\, .
\ee
Finally we denote by $Q_B$ the BRST charge and by $\XX(z)=\{Q_B,\xi(z)\}$ the
picture changing operator.

Since our construction will make use of some of the results of \cite{1408.0571}, we
review here the pertinent results. We denote by
$\MM_{g,n}$ the moduli space of Riemann surfaces of genus $g$ with $n$
distinguishable punctures.
$\wt\PP_{g,n}$ will denote a fiber bundle over $\MM_{g,n}$ where the fiber is
infinite dimensional, containing information about a choice of local coordinate 
system (up to phases) around each puncture and locations of $2g+n-2$ PCO's 
on the Riemann surface.\footnote{This is a generalization of corresponding construction
in bosonic string theory\cite{nelson,9206084} where the choice of local coordinate system is the
only data on the fiber.}
In \cite{1408.0571} we constructed, for a set of 
states $\{|\Phi_1\rangle,\cdots |\Phi_n\rangle\}\in\HH_0$,
a $p$-form 
$\Omega^{(g,n)}_p(|\Phi_1\rangle, \cdots |\Phi_n\rangle)$ on $\wt\PP_{g,n}$
for arbitrary $p$, satisfying
\be \label{eomega}
\sum_{i=1}^n (-1)^{n_1+\cdots n_{i-1}} \Omega^{(g,n)}_p(|\Phi_1\rangle, \cdots |\Phi_{i-1}\rangle,
Q_B|\Phi_i\rangle, |\Phi_{i+1}\rangle, \cdots |\Phi_n\rangle)=(-1)^p d\Omega^{(g,n)}_{p-1}
(|\Phi_1\rangle,\cdots, |\Phi_n\rangle) \, ,
\ee
where $d$ denotes exterior derivative on $\wt\PP_{g,n}$ and $n_i$ is the ghost number
of $|\Phi_i\rangle$.
$\Omega^{(g,n)}_p(|\Phi_1\rangle, \cdots |\Phi_n\rangle)$ 
is a multilinear function of the $n$ states $|\Phi_1\rangle, \cdots |\Phi_n\rangle$,
constructed in terms of
a correlation function on the Riemann surface with the state $|\Phi_i\rangle$ inserted at
the $i$-th puncture using the local coordinate system at that puncture.
It has the symmetry property
\be \label{exchange}
s_{i,i+1}\circ \Omega^{(g,n)}_p(|\Phi_1\rangle, \cdots |\Phi_{i-1}\rangle,
|\Phi_{i+1}\rangle, |\Phi_{i}\rangle,|\Phi_{i+2}\rangle\cdots
 |\Phi_n\rangle)= (-1)^{n_i n_{i+1}} \, \Omega^{(g,n)}_p(|\Phi_1\rangle, \cdots |\Phi_n\rangle)
\ee
where $s_{i,i+1}$ is the transformation on $\wt\PP_{g,n}$ that exchanges the punctures
$i$ and $i+1$ together with their local coordinates
and $s_{i,i+1}\circ \Omega^{(g,n)}_p$ is the pullback of
$\Omega^{(g,n)}_p$ under this transformation.
Another useful property of $\Omega^{(g,n)}_p(|\Phi_1\rangle, \cdots |\Phi_n\rangle)$
is that it is non-vanishing only if the ghost numbers of
$|\Phi_1\rangle, \cdots |\Phi_n\rangle$ add up to $p-6g+6$.

The analysis of \cite{1311.1257,1401.7014,1408.0571} 
also identified, for each $g,n$, a $6g-6+2n$
dimensional subspace $\KK_{g,n}$
of $\MM_{g,n}$ called the 1PI region and
a `section' $\RR_{g,n}$ of $\wt\PP_{g,n}$ over this subspace with the following
properties\footnote{In \cite{1311.1257,1401.7014,1408.0571} we did not use 
the specific symbols
$\KK_{g,n}$ and $\RR_{g,n}$ for these subspaces. This notation is introduced
in  this paper. Also, while the concrete algorithm for constructing these regions was
given in \cite{1311.1257,1401.7014,1408.0571}, 
explicit construction of these regions was not given as there is
a lot of freedom. A concrete example
can be provided by  computing the off-shell 1PI amplitudes in 
the closed bosonic string field theory of \cite{9206084}
in Siegel gauge. 
The subspace of the
moduli space covered by the 1PI amplitudes will give $\KK_{g,n}$ and the local coordinates
at the punctures induced from this construction will give $\RR_{g,n}$ sans the
information on the PCO's. The PCO locations will then have to be chosen consistent
with gluing compatibility and avoiding spurious singularities using the trick of vertical
integration described in \cite{1408.0571}.}
\begin{enumerate}
\item For some $(g,n)$,  $\RR_{g,n}$
could have `vertical segments' along which the locations of the PCO's change without
any change in the moduli parametrizing the base. This is necessary for avoiding the
spurious singularities which occur in type II / heterotic string perturbation 
theory\cite{Verlinde:1987sd}. The
rules for carrying out the integration along these vertical segments have been 
described in \cite{1408.0571} and further refined in \cite{unpub}. For this reason 
we shall refer to $\RR_{g,n}$ as an integration cycle instead of a section.
\item Even though the punctures are distinguishable, we choose
$\RR_{g,n}$ to be symmetric under the exchange of the punctures. This means
that for every point in $\RR_{g,n}$ we'll also have in $\RR_{g,n}$
the point obtained by exchanging
any two punctures together with their local coordinate systems, and the locations of
the PCO's will remain unchanged  under this exchange.
\item Some time we may encounter a situation in which a single integration cycle cannot be chosen
consistent with this symmetry principle. A simple example is three punctured sphere
requiring one insertion of the picture changing operator. In order to be consistent with
the symmetry principle we would require that the location of the PCO will be invariant
under any $SL(2,C)$ transformation that permutes these three punctures. It is easy to
verify that there is no such point on the sphere. However we can choose a pair of
points which are permuted among themselves under the $SL(2,C)$ transformations
that permute the three
punctures.  Thus we can restore full permutation symmetry by taking the average
of these two choices for the location of the PCO. To deal with such situations we allow
$\RR_{g,n}$ to be formal weighted average of multiple integration cycles. Since eventually we
shall be interested in integrating $\Omega^{(g,n)}_{6g-6+2n}$ 
over these integration cycles, the integral over the
weighted average of integration cycles can be regarded as the weighed average of the integrals 
over the corresponding integration cycles.

\item Take a Riemann surface (equipped with choice of local coordinate system and
arrangement of PCO's) corresponding to a point in $\RR_{g_1,n_1}$
and another Riemann surface corresponding to a point in $\RR_{g_2,n_2}$.
If we take one puncture from each of these Riemann surfaces, denote the local
coordinates around these punctures
by $z$ and $w$ with the punctures being located at $z=0$ and $w=0$, and
glue the two Riemann surfaces by the plumbing fixture relation
\be \label{eplumb}
z\, w = e^{-s+i\theta}\, , \qquad 0\le s<\infty, \quad 0\le\theta<2\pi,
\ee
we get a two parameter family of Riemann surfaces of genus $g_1+g_2$ and
$n_1+n_2-2$ punctures, labelled by $s$ and $\theta$. 
Considering the $6g_i-6+2n_i$
parameter family of Riemann surfaces contained in $\RR_{g_i,n_i}$ we get
altogether $6(g_1+g_2) + 2(n_1+n_2) - 10$ parameter family of Riemann 
surfaces.
These describe a subspace of $\MM_{g_1+g_2, n_1+n_2-2}$ which we shall denote by
$\KK_{g_1,n_1}\circ \KK_{g_2,n_2}$. 
Furthermore the choice of local coordinates at the punctures and the locations
of PCO's on the original surfaces automatically produce similar data on the
final Riemann surfaces. Thus we really have a $6(g_1+g_2) + 2(n_1+n_2) - 10$ 
dimensional 
subspace of $\wt\PP_{g_1+g_2,n_1+n_2-2}$.
We shall call this $\RR_{g_1,n_1}\circ \RR_{g_2,n_2}$.
Summing over all
inequivalent permutations of the external punctures we generate a subspace of
$\wt\PP_{g_1+g_2, n_1+n_2-2}$ that we shall call 
${\bf S}[\RR_{g_1,n_1}\circ \RR_{g_2,n_2}]$.

\item Similarly by gluing three families of 
Riemann surfaces associated with $\RR_{g_1,n_1}$, $\RR_{g_2,n_2}$ and
$\RR_{g_3,n_3}$ in all possible ways, but using only two plumbing fixtures so
that no closed loop is formed, we generate a new 
$6(g_1+g_2+g_3)+2(n_1+n_2+n_3)-14$ parameter
family of Riemann surfaces
of genus $g_1+g_2+g_3$ and $n_1+n_2+n_3-4$ punctures. These correspond
to a subspace of $\MM_{g_1+g_2+g_3,n_1+n_2+n_3-4}$. Again the choice of
the local coordinates and the PCO data on the original Riemann surfaces generate
similar data on the final Riemann surfaces, producing a subspace of $\wt\PP_{g_1+g_2+g_3,
n_1+n_2+n_3-4}$. This process can be continued, generating more and more families of
Riemann surfaces by gluing more and more 1PI families of Riemann surfaces in 
all possible inequivalent ways.
\item Now consider a given $g$ and $n$. Following the procedure described above 
we can generate a large class of Riemann surfaces of genus $g$ with $n$ punctures,
-- 
some from 1PI region
$\KK_{g,n}$ of the moduli space, some from gluing two 1PI family of Riemann surfaces
of lower genera and/or lower number of punctures and so on. The claim is that this
exhausts the whole of $\MM_{g,n}$ in a one to one fashion, -- indeed $\KK_{g,n}$ is
chosen so as to precisely account for any deficit that we might have from gluing 
surfaces of lower
genera / lower number of punctures. Furthermore the gluing compatible choice of
integration cycles guarantee that the choice of the integration cycle over $\MM_{g,n}$
that we get this way is continuous across the boundaries between different types of
contributions {\it e.g.} the 
boundary of $\RR_{g_1,n_1}\circ
\RR_{g-g_1, n-n_1+2}$ from the $s=0$ end of the plumbing fixture relation
\refb{eplumb} smoothly
matches a component of the boundary of 
$\RR_{g,n}$. Again, the construction of $\RR_{g,n}$ is designed to ensure this by choosing the
integration cycle $\RR_{g,n}$ on $\KK_{g,n}$ such that it smoothly matches the integration cycle 
$\RR_{g_1,n_1}\circ \RR_{g-g_1, n-n_1+2}$ on $\KK_{g_1,n_1}\circ \KK_{g-g_1, n-n_1+2}$
at the common boundary of $\KK_{g,n}$ and $\KK_{g_1,n_1}\circ \KK_{g-g_1, n-n_1+2}$.
We shall 
denote the integration cycle of $\wt\PP_{g,n}$ over the whole of
$\MM_{g,n}$ obtained this way by $\SSS_{g,n}$. Also all parts of 
$\SSS_{g,n}$ outside $\RR_{g,n}$ will be called one particle reducible (1PR)
part of $\SSS_{g,n}$ 
and the associated Riemann surfaces will be called 1PR Riemann
surfaces.

\item In this description of $\MM_{g,n}$, all separating type degenerations -- where a
Riemann surface degenerates into two or more distinct Riemann surfaces joined
by a long handle -- comes from 
1PR Riemann surfaces with the
parameter $s$ in \refb{eplumb} approaching infinity for one or more of the plumbing
fixture. Thus the 1PI family 
$\RR_{g,n}$ does not contain any boundary associated with separating type 
degeneration. However $\RR_{g,n}$ can (and does) contain boundaries corresponding to
non-separating type degenerations where two punctures on the same Riemann
surface are connected by a long handle.  
\end{enumerate}

Once these different regions have been identified, we define the genus $g$, $n$-point
off-shell amplitude with external states $|\Phi_1\rangle,\cdots |\Phi_n\rangle$ to be the
integral of $\Omega^{(g,n)}_{6g-6+2n}(|\Phi_1\rangle,\cdots |\Phi_n\rangle)$ over the
subspace $\SSS_{g,n}$. The contribution to this integral from the 1PI part
$\RR_{g,n}$ of $\SSS_{g,n}$
will be called the 1PI contribution to the off-shell amplitude, while the contribution from the 1PR
part of $\SSS_{g,n}$ 
will be called the 1PR contribution. This will be justified later when we construct the
1PI effective action for the string field theory and compute the off-shell amplitude from this
effective action. We should note here that off-shell amplitudes refer to off-shell
Green's functions with external tree level propagators truncated.

Readers familiar with the construction of the covariant string field theory\cite{9206084}
would recognize the close resemblance between the regions $\RR_{g,n}$ used here
and the
regions $\VV_{g,n}$ used in the construction of the quantum master action of 
bosonic string field
theory. Of course one difference is that in the construction of \cite{9206084} there was
no information about the PCO's, so the regions $\VV_{g,n}$ were subspaces of $\wh\PP_{g,n}$
which had $\MM_{g,n}$ as the base and the data on local coordinates (modulo 
phases) at the punctures as
fibers. The main difference however is that while building the analog of
the integration cycle $\SSS_{g,n}$ over the
whole of $\MM_{g,n}$ from the plumbing fixture of the Riemann surfaces associated with
$\VV_{g,n}$, we also allow closed loops. Thus for example we can glue two punctures on
the same Riemann surface associated with $\VV_{g,n}$, glue two punctures of
a Riemann surface associated with $\VV_{g_1,n_1}$ with two punctures of a Riemann
surface associated with $\VV_{g_2,n_2}$ and so on. As a result all degenerations --
separating as well as non-separating type -- come from one or more of the plumbing
fixture parameters $s$ approaching infinity, and $\VV_{g,n}$ never
contains any degenerate Riemann surface. If we take the $\VV_{g,n}$'s of \cite{9206084}
and glue them together in all possible ways using $\refb{eplumb}$ including configuration
with loops, but not allowing
those configurations where by taking  the parameter $s$ associated with
{\it one of the plumbing fixture relations} to infinity 
we reach a separating type degeneration,
we get possible candidates for $\RR_{g,n}$ (without the PCO data). 
This is precisely analogous to the construction
of 1PI amplitudes using Feynman diagrams of a quantum field theory.

The boundary of $\RR_{g,n}$ is an object of special interest since the integral of a 
total derivative on $\RR_{g,n}$ will receive contribution from this boundary. 
As already discussed, 
$\RR_{g,n}$ has boundaries corresponding to non-separating type degenerations.
However they will not be of interest to us since the boundary contributions from these
boundaries can be made to vanish using appropriate $i\eps$ 
prescription\cite{berera,1307.5124}. Thus from now on we shall ignore the existence of these
boundaries. The
other boundaries of $\RR_{g,n}$ coincide with the boundaries of 
$\RR_{g_1,n_1}\circ \RR_{g-g_1,n-n_1+2}$ arising from the $s=0$ end of the
plumbing fixture relation \refb{eplumb} since $\RR_{g,n}$ and 
$\RR_{g_1,n_1}\circ \RR_{g-g_1,n-n_1+2}$ will have to `fit together' inside $\wt\PP_{g,n}$
to generate part of the smooth integration cycle $\SSS_{g,n}$.
Denoting this boundary by $\{\RR_{g_1,n_1},
\RR_{g_2,n_2}\}$ we get
\be  \label{eboundary}
\p \RR_{g,n} = -{1\over 2} \sum_{g_1,g_2\atop g_1+g_2=g} \sum_{n_1,n_2\atop n_1+n_2 = n+2}
{\bf S}[\{\RR_{g_1,n_1} , \RR_{g_2,n_2}\}]\, ,
\ee
where, as before, 
${\bf S}$ denotes the operation of summing over all inequivalent permutation
of the external punctures. 
The factor of $1/2$ reflects that on the right hand the same term appears twice due
to the exchange symmetry $g_1\leftrightarrow g_2$, $n_1\leftrightarrow n_2$.
Even for $g_1=g_2$ and $n_1=n_2$ there is a double counting 
since ${\bf S}$ sums over
exchanges of the puncture labels. The
$-$ sign reflects that the boundaries of $\RR_{g,n}$ and 
$\RR_{g_1,n_1}\circ \RR_{g_2,n_2}$ must be oppositely oriented since 
$\RR_{g,n}$ together with $(1/2) \sum_{g_1,g_2\atop g_1+g_2=g} 
\sum_{n_1,n_2\atop n_1+n_2 = n+2}
\RR_{g_1,n_1} \circ \RR_{g_2,n_2}$ fill part of $\SSS_{g,n}$.  Physically
$\{\RR_{g_1,n_1} , \RR_{g_2,n_2}\}$ represents the set of punctured Riemann
surfaces equipped with choice of local coordinates at the punctures and PCO
locations that we obtain by gluing the families of Riemann surfaces  corresponding
to $\RR_{g_1.n_1}$ and $\RR_{g_2,n_2}$ using plumbing fixture relation \refb{eplumb}
with the parameter
$s$ set to zero. The orientation of $\{A,B\}$ will be defined by taking its volume form to be
$d\theta\wedge dV_A\wedge dV_B$ where $dV_A$ and $dV_B$ are volume forms on 
$A$ and $B$ respectively.

It follows from general properties of conformal field theories on Riemann surfaces that
on $\{\RR_{g_1,n_1} , \RR_{g_2,n_2}\}$, $\Omega^{(g_1+g_2,n_1+n_2-2)}_p$
satisfies the factorization property
\ben \label{efactor}
&&\int_\theta \, \Omega^{(g_1+g_2,n_1+n_2-2)}_p(|\Phi_1\rangle, \cdots 
|\Phi_{n_1+n_2-2}\rangle)
\nonumber \\
&=& \sum_{p_1,p_2\atop p_1+p_2 = p-1} (-1)^{n_1+\cdots n_{n_1-1}+p_1+p_2+p_1p_2} 
\, \Omega^{(g_1,n_1)}_{p_1}(|\Phi_1\rangle,\cdots |\Phi_{n_1-1}\rangle, |\vp_r\rangle)
\nonumber \\
&& \qquad \qquad \qquad
\wedge\,  \, \Omega^{(g_2,n_2)}_{p_2}( |\vp^r\rangle, |\Phi_{n_1}\rangle,\cdots |\Phi_{n_2-2}\rangle)
\een
where $\int_\theta$ denotes the result of integration over the angular coordinate
$\theta$ appearing in the plumbing fixture relation \refb{eplumb},
and $\{|\vp_r\rangle\}$ and
$\{|\vp^r\rangle\}$ are a set of dual basis of $\HH_0$ satisfying
\be \label{ebasis}
\langle \vp^r | c_0^-|\vp_s\rangle = \delta^r{}_s \qquad \Leftrightarrow \qquad
\langle \vp_s | c_0^-|\vp^r\rangle = \delta^r{}_s  \, .
\ee
In arriving at \refb{efactor} one has to use the fact that the $\theta$ integral produces
the projector $\delta_{L_0,\bar L_0}$ and is accompanied by an insertion of
$b_0^-$ in the correlator. Together they ensure that the sum over the complete set of
states $|\vp_r\rangle$, $|\vp^r\rangle$ run only over states in $\HH_0$ satisfying 
\refb{econd}. The $(-1)^{n_1+\cdots n_{n_1-1}+p_1+p_2+p_1p_2}$ 
factor can be obtained from the results of \cite{1408.0571} after taking into account an extra
minus sign due to the fact that here we are picking the boundary contribution from
the lower limit of the $s$ integral while \cite{1408.0571} analyzed the boundary contribution from
the upper limit. We also need to use the fact that the basis states $\langle\vp_r^c|$ 
of \cite{1408.0571} are
related to $\langle\vp^r|$ given here by $\langle\vp_r^c|=\langle\vp^r| c_0^-$.

\sectiono{The 1PI effective action} \label{saction}

The construction of the 1PI action of NS sector fields 
will be a generalization of the tree level action for NS sector fields given in
\cite{9202087}. Most of the properties of this theory 
discussed in this section will involve adapting the
various results for the classical bosonic string field theory of \cite{saadi,kugo} to
the 1PI effective action of heterotic / type II string field theory. However we shall
try to keep the discussion self-contained by briefly outlining the proofs of the various 
results we shall describe.

We shall take the string field to be an element of the small Hilbert 
 space as given by the $\eta_0|s\rangle=0$ condition 
 in \refb{econd}\cite{FMS}. 
 This was the spirit of the construction in \cite{wittenssft,9202087}, and
 some of the recent approaches, {\it e.g.} the ones  
 considered in \cite{1312.2948,1403.0940} also falls broadly in this class.
There are also alternate approaches based on taking the string field
to be an element of the large Hilbert space, {\it e.g.} the ones used in
\cite{9503099,0109100,0406212,0409018,0911.2962} -- but at present
we do not know whether this approach can be generalized to give a fully
consistent 1PI effective action for the NS sector fields.

\subsection{The $\{~\}$ and $[~]$ products} 

We  define, for $|\Phi_i\rangle\in\HH_0$, 
\be \label{edefcurly}
\{\Phi_1\cdots \Phi_n\}= \sum_{g=0}^\infty (g_s)^{2g} 
\int_{\RR_{g,n}} \Omega^{(g,n)}_{6g-6+2n}(|\Phi_1\rangle,\cdots |\Phi_n\rangle)\, .
\ee
If $\RR_{g,n}$
contains vertical segments then one has to use the procedure described in 
\cite{1408.0571,unpub}
for carrying out the integral over these segments.
We also define $|[\Phi_2\cdots \Phi_n]\rangle\in\HH_0$ via the relations
\be \label{edefsquare}
\langle \Phi_1| c_0^- |[\Phi_2\cdots \Phi_n]\rangle = \{\Phi_1\cdots \Phi_n\}
\ee
for all $|\Phi_1\rangle\in\HH_0$. Here $\langle A|B\rangle$ denotes the BPZ inner product.
Note that to keep the notation simple we have dropped the ket symbol
$|~\rangle$ from the states when they appear in the argument of
$\{~\}$ or $[~]$. Also we shall drop the ket symbol from $|[\Phi_2\cdots \Phi_n]\rangle$
except in inner products, i.e.\  $[\Phi_2\cdots \Phi_n]$ will actually denote
$|[\Phi_2\cdots \Phi_n]\rangle$. It follows from the property of $\Omega^{(g,n)}_p$
that in order to get non-vanishing result for $\{\Phi_1\cdots \Phi_n\}$ 
we must have $\sum_{i=1}^n n_i = 2n$ 
where $n_i$ is the ghost number of $\Phi_i$, and that 
$[\Phi_2\cdots \Phi_n]$ has ghost number equal to $3 +\sum_{i=2}^n n_i-2(n-1)$.

Based on the identities \refb{eomega} and \refb{eboundary} one can now establish
a set of identities involving $Q_B$ and $[~]$. Since their derivation is identical to the
ones for bosonic string field theory\cite{9206084} we shall be brief.
The first
identity gives the symmetry properties of $[\cdots]$ and $\{\cdots\}$:
\ben \label{esymmetry}
&& [\Phi_2\cdots \Phi_{i-1}\Phi_{i+1} \Phi_i\Phi_{i+2} \cdots \Phi_n]
=(-1)^{n_i n_{i+1}} [\Phi_2\cdots \Phi_n]\nonumber \\
&& \{\Phi_1 \Phi_2\cdots \Phi_{i-1}\Phi_{i+1} \Phi_i\Phi_{i+2} \cdots \Phi_n\}
=(-1)^{n_i n_{i+1}} \{\Phi_1\Phi_2\cdots \Phi_n\}\, .
\een
The inner product of the first equation of \refb{esymmetry} with an arbitrary bra
$\langle \Phi_1|c_0^-$, which is also the second equation of \refb{esymmetry}, 
comes from the
integral of \refb{exchange} over $\RR_{g,n}$.
The second useful identity is
\be \label{eneweq}
\{\Phi_1\cdots \Phi_k[\wt\Phi_1\cdots \wt\Phi_\ell]\}
= -\{\wt\Phi_1\cdots \wt\Phi_\ell[\Phi_1\cdots \Phi_k]\} \quad \hbox{for $|\Phi_i\rangle, |\wt\Phi_j\rangle
\in\HH_0$}\, .
\ee
To prove this we note that as a consequence of \refb{ebasis}, $\sum_r |\vp_r\rangle 
\langle \vp^r|c_0^-$ and $\sum_r |\vp^r\rangle 
\langle \vp_r|c_0^-$ act as identity operator on states in $\HH_0$. 
If we insert the first operator in front of $[\cdots]$ on the left hand side
of \refb{eneweq} and the second operator  
in front of $[\cdots]$ on the right hand side
of \refb{eneweq}, we see that the two  expressions are identical except for a sign and
the
locations of $\vp_r$ and $\vp^r$ in the correlation function. Using the ghost number
conservation laws mentioned below \refb{edefsquare}, the fact that ghost numbers
of $\vp_r$ and
$\vp^r$ are related by $n_{\vp_r}=5 - n_{\vp^r}$, and \refb{esymmetry}
we can show that the 
arguments inside $\{\cdots \}$ on the two sides can be brought to the same arrangement
at the cost of
generating an extra minus sign. This cancels the explicit minus sign in \refb{eneweq}
showing that the two sides are equal.

The third equation, known as the main
identity\cite{9206084}, tells us that for $n\ge 1$,\footnote{In 
the analysis of \cite{9206084} there is an additional contribution
involving $\Delta$ operation in \refb{eboundary} which gives rise to additional terms in
the identities involving $Q_B$ and $[~]$. These terms are absent here since we build up
the moduli space $\MM_{g,n}$ from 1PI Riemann surfaces instead of the Riemann
surfaces associated with the elementary vertices of string field theory.} 
\ben \label{emain}
&& Q_B[\Phi_2\cdots \Phi_n] + \sum_{i=2}^n (-1)^{n_2+\cdots n_{i-1}}[\Phi_2\cdots \Phi_{i-1} Q_B \Phi_i
\Phi_{i+1} \cdots \Phi_n] \nonumber \\
&& +  \sum_{\ell,k\ge 0\atop \ell+k=n-1} \sum_{\{i_r;r=1,\cdots \ell\}, \{j_s;s=1,\cdots k\}\atop
\{i_r\}\cup \{j_s\} = \{2,\cdots n\}
}\sigma(\{i_r\}, \{j_s\})
[\Phi_{i_1} \cdots \Phi_{i_\ell} [\Phi_{j_1} \cdots \Phi_{j_k}]] = 0
\een
where in
the last term the sum runs over all possible ways of splitting the
set $\{2,\cdots n\}$ into the set $\{i_r\}$ and the set $\{j_s\}$. 
$\sigma(\{i_r\}, \{j_s\})$ is the sign that one picks up while rearranging
$Q_B,\Phi_2,\cdots \Phi_n$ to 
$\Phi_{i_1},\cdots \Phi_{i_\ell}, Q_B, \Phi_{j_1},\cdots \Phi_{j_k}$.
The inner product of \refb{emain} with an arbitrary bra
$\langle\Phi_1|c_0^-$ is obtained by integrating
\refb{eomega} over $\RR_{g,n}$ and multiplying both sides by $(-1)^{n_1}$. 
The first two terms in \refb{emain}
come from the integral of
the left hand side of
\refb{eomega} whereas the last term has its origin in the integral of the
right hand side of
\refb{eomega}. The latter, being a total derivative, 
receives contribution from the boundary $\p\RR_{g,n}$
described in \refb{eboundary}. This can then be evaluated using \refb{efactor}.
The terms that are generated are of two types: $\{\Phi_1 \Phi_{i_1}\cdots \Phi_{i_\ell}
[\Phi_{j_1}\cdots \Phi_{j_k}]\}$ and $\{\Phi_{j_1}\cdots \Phi_{j_k}[\Phi_1 
\Phi_{i_1}\cdots \Phi_{i_\ell}]\}$.
Using \refb{eneweq} we can show that the two terms are in
fact identical after performing the sum over the sets $\{i_r\}$ and $\{j_s\}$.
This allows us to keep only the first term and cancel the extra factor of 1/2 in 
\refb{efactor}. This can then be interpreted as the inner product of
$-\langle \Phi_1|c_0^-$ with the last term on the left hand side of
\refb{emain}.

Note that inside $[\cdots]$ in the first term of \refb{emain}
the first argument is $\Phi_2$ and hence there are
only $n-1$ arguments. Thus for $n=1$ we have the equation $Q_B[] +[[]]=0$.

\subsection{The action and its gauge invariance}

We are now ready to describe the
1PI effective action and its gauge invariance. The string field $|\Psi\rangle$ is
taken to be an element of $\HH_0$ of ghost number 2. The action is given by
\be \label{eaction}
S(|\Psi\rangle) = g_s{}^{-2} \left[{1\over 2}\,  
\langle \Psi| c_0^- Q_B|\Psi\rangle + \sum_{n=1}^\infty {1\over n!}
 \{ \Psi^n \}\right]
\ee
where $\{\Psi^n\}$ denotes $\{\Psi \Psi \cdots \Psi\}$ with $n$ copies of
$\Psi$ inside $\{ ~ \}$.
The gauge transformation law for $|\Psi\rangle$ is given by
\be  \label{egauge}
\delta|\Psi\rangle = Q_B|\Lambda\rangle + \sum_{n=0}^\infty {1\over n!} 
[\Psi^n \Lambda]
\ee
where $|\Lambda\rangle$ is an element of $\HH_0$ with ghost number 1. 
Proof of gauge invariance of the action \refb{eaction} proceeds in the same
way as in \cite{saadi,kugo,9206084} except that in classical string field theory
the range of summation over $n$ in \refb{eaction} and \refb{egauge} begins
at 3 and 1 respectively. We shall now briefly discuss the proof.
Under the transformation \refb{egauge} the change $\delta S$ in the action is given by
\be \label{exx1}
g_s{}^2\, \delta S = \sum_{n=0}^\infty {1\over n!} \langle\Psi |c_0^- Q_B
[\Psi^n\Lambda]\rangle + \sum_{n=1}^\infty {1\over (n-1)!} \{\Psi^{n-1}Q_B\Lambda\}
+ \sum_{n=1}^\infty {1\over (n-1)!} \sum_{m=0}^\infty {1\over m!} \{\Psi^{n-1} 
[\Psi^m\Lambda]\}
\, ,
\ee
where we have used $Q_B{}^2=0$ to set one of the terms to 0.
Now the first two terms on the right hand side can be manipulated as
\ben \label{exx2}
&& \sum_{n=0}^\infty {1\over n!} \langle Q_B\Psi |c_0^- 
[\Psi^n\Lambda]\rangle 
+  \sum_{n=1}^\infty {1\over (n-1)!} \langle Q_B\Lambda|c_0^-| [\Psi^{n-1}]\rangle \nonumber \\
&=& \sum_{n=0}^\infty {1\over n!} \{Q_B\Psi\Psi^n\Lambda\}
-  \sum_{n=1}^\infty {1\over (n-1)!} \langle \Lambda|c_0^- Q_B[\Psi^{n-1}]\rangle
\nonumber \\
&=& -\sum_{n=0}^\infty {1\over n!} \langle \Lambda|c_0^-|[Q_B\Psi\Psi^n]\rangle
-  \sum_{n=1}^\infty {1\over (n-1)!} \langle \Lambda|c_0^- Q_B[\Psi^{n-1}]\rangle\, .
\een
We now interpret the first term on the right hand side as the average of $n+1$ terms
where $Q_B$ acts on each of the $\Psi$'s inside $[\cdots ]$ once, make change of
variables  $n=m-1$ in the first term and $n=m+1$ in the second term
and
use \refb{emain} with all the $\Phi_i$'s equal to $\Psi$ to express the right hand
side of \refb{exx2} as
\be
 \sum_{m=0}^\infty {1\over m!} \sum_{0\le n_1,n_2\le \infty\atop n_1+n_2=m}
{m!\over n_1! n_2!}
\langle \Lambda|c_0^- |  [\Psi^{n_1}[\Psi^{n_2}]]\, .
\ee
The $m!/(n_1! n_2!)$ term comes from the number of ways we can divide the
$m$ objects into $n_1$ objects and $n_2=m-n_1$ objects.
Regarding $n_1,n_2$ as independent summation variables we can now express this as
\be 
\sum_{n_1,n_2=0}^\infty {1\over n_1! n_2!} \{ \Lambda \Psi^{n_1}[\Psi^{n_2}]\}
= -\sum_{n_1,n_2=0}^\infty {1\over n_1! n_2!}  \{ \Psi^{n_2}[\Lambda \Psi^{n_1}]\}
\ee
where we have used \refb{eneweq}. Identifying $(n_1,n_2)$ with  $(m,n-1)$ we see that
this exactly cancels the third term in \refb{exx1}. This establishes gauge invariance of the
action.

It is also easy to verify that if we use the Siegel gauge fixing $(b_0+\bar b_0)|\Psi\rangle=0$,
then the tree level amplitude computed from this action produces the full off-shell amplitude
given by
\be
\sum_g g_s{}^{2g-2}\, \int_{\SSS_{g,n}}  
\Omega^{(g,n)}_{6g-6+2n}(|\Phi_1\rangle,\cdots |\Phi_n\rangle)\, .
\ee
Essentially joining two or more interaction vertices by a Siegel gauge propagator
corresponds to joining the corresponding Riemann surfaces by the plumbing
fixture relation \refb{eplumb}.
The sum of all tree level graphs then produces the union of all 
integration cycles
that can be obtained either from 1PI surfaces of genus $g$ with $n$ punctures, or by
gluing 1PI surfaces of lower genera / lower number of punctures
using plumbing fixture to produce a genus $g$
Riemann surface with $n$ punctures. By construction this produces all of $\SSS_{g,n}$.

Note that if we relax the condition that $|\Psi\rangle$ carries ghost number 2 and
allow it to carry arbitrary ghost number, we shall obtain the Batalin-Vilkovisky (BV)
master action\cite{9206084}. However this will satisfy the classical master equation instead of
the quantum master equation. This is in consonance with the fact that we are
supposed to compute only tree amplitudes using this action. However since in
computation of tree diagrams only ghost number two states appear even as
intermediate states, the extra fields in this master action do not seem to play any
useful role here.

\subsection{Equations of motion and the vacuum} \label{esolution}

The equations of motion derived from the action are given by
\be
Q_B|\Psi\rangle + \sum_{n=1}^\infty {1\over (n-1)!} [\Psi^{n-1}]=0\, .
\ee
Note the constant term arising from the $n=1$ term. This is related to the presence
of the 
linear term proportional to $\{\Psi\}$ in the 1PI
effective action from one loop onwards. Thus $|\Psi\rangle=0$ is not  a
classical solution, and we have to solve the classical equations of motion derived from
the 1PI action to find the correct ground state. 
This can be done by looking for a
solution to these equations iteratively as a power series in $g_s$, starting
with $|\Psi\rangle =0$ as the leading solution at order $g_s{}^0$. Thus if $|\Psi_k\rangle$
denotes the solution to order $g_s{}^k$ then we can 
write the equation as\footnote{Even though the natural loop
expansion parameter is $g_s{}^2$, in some special cases, {\it e.g.} the one discussed in
\S\ref{sr.3}, the classical solution $|\Psi\rangle$ begins its expansion at order $g_s$.
For this reason we have taken $g_s$ as the expansion parameter.} 
\be \label{etest0}
Q_B|\Psi_{k+1}\rangle = -\sum_{n=1}^\infty {1\over (n-1)!} [\Psi_k^{n-1}] + \OO(g_s{}^{k+2})\, ,
\ee
and solve this iteratively. Since the expansion of $|\Psi_k\rangle$ begins at order $g_s$,
the sum over $n$ on the right hand side can be truncated at $n\le k+2$, but we
shall keep the upper limit to be $\infty$ for the ease of manipulating the terms. 
In order to ensure that there is no obstruction to this procedure we have to ensure
that the right hand side is a BRST trivial state to order $g_s{}^{k+1}$. 
A necessary condition for this is that the right hand side is BRST invariant, i.e.
\be \label{etest1}
Q_B \sum_{n=1}^\infty {1\over (n-1)!} [\Psi_k^{n-1}]\
\ee
vanishes to order $g_s{}^{k+1}$. 
This is known to be true since the
early days of string field theory (see {\it e.g.} \cite{mukherji}), 
but we shall briefly describe the proof for
completeness. 
Using \refb{emain} we can express \refb{etest1} as
\be \label{e3.11}
- \sum_{n=2}^\infty {1\over (n-2)!} [ \Psi_k^{n-2}(Q_B\Psi_k)]
- \sum_{n=1}^\infty {1\over (n-1)!} \, \sum_{n_1,n_2\ge 0\atop n_1+n_2=n-1}
{(n-1)!\over n_1! n_2!} [\Psi_k^{n_1}[\Psi_k^{n_2}]]\, .
\ee
In the first term we can substitute for $Q_B\Psi_k$ using \refb{etest0} 
with $k$ replaced by $k-1$ on the left hand side, but  
on the right hand side we can continue to use $\Psi_k$ since 
the net error we shall make by this in \refb{e3.11} will be of order $g_s^{k+2}$.
The second term can be
simplified by taking $n_1$ and $n_2$ to be independent summation variables. This
reduces \refb{e3.11} to
\be
\sum_{n=2}^\infty {1\over (n-2)!} \sum_{m=1}^\infty {1\over (m-1)!}
[\Psi_k^{n-2}[\Psi_k^{m-1}]]
- \sum_{n_1,n_2\ge 0} {1\over n_1!n_2!} \, [\Psi_k^{n_1}[\Psi_k^{n_2}]]
+\OO(g_s{}^{k+2})\, .
\ee
After identifying $n_1$ with $n-2$ and $n_2$ with $m-1$ these two terms cancel showing
that \refb{etest1} vanishes to order $g_s{}^{k+1}$. 

This in turn
allows us to write down an explicit solution of \refb{etest0} in the $L_0^+\ne 0$ sector.
Denoting by ${\bf P}$ the projection operator into zero momentum
$L_0^+=0$ states and
using $\{Q_B, b_0^+\}=L_0^+$, we can write
the iterative solution to \refb{etest0} as
\be \label{esoln}
|\Psi_{k+1}\rangle = -{b_0^+\over L_0^+} \sum_{n=1}^\infty {1\over (n-1)!} (1-{\bf P})
[\Psi^{n-1}]
+ Q_B|s_{k+1}\rangle + |\Phi_{k+1}\rangle\, ,
\ee
where $|s_{k+1}\rangle$ denotes the expansion of some arbitrary ghost number
one, zero
momentum state $|s\rangle\in\HH_0$
to order $g_s{}^{k+1}$ and $|\Phi_{k+1}\rangle$ satisfies
\be \label{esol1}
{\bf P}|\Phi_{k+1}\rangle = |\Phi_{k+1}\rangle, \qquad Q_B|\Phi_{k+1}\rangle
= - \sum_{n=1}^\infty {1\over (n-1)!} {\bf P}
[\Psi_k^{n-1}]+\OO(g_s{}^{k+2})\, .
\ee
The existence of the solutions to the second equation is not automatic. 
To see this let us suppose that 
$|V_i\rangle$ for $i=1,\cdots N$ is a basis of
states in $\HH_0$ describing non-trivial elements of the BRST cohomology and carrying
zero momentum, ghost number 2 and $L_0^+=0$. These in fact represent
zero momentum massless states. Then in order for the second equation in
\refb{esol1} to have solution, we need
\be \label{eobstruct}
\langle V_i|c_0^-| \sum_{n=1}^\infty {1\over (n-1)!} [\Psi_k^{n-1}]\rangle=
\OO(g_s{}^{k+2}) \quad \hbox{for
$1\le i\le N$} \, .
\ee
Unless this holds at each order in $g_s$,
we cannot solve \refb{etest0} by the iterative
procedure we have described. As we shall discuss shortly, this will be
related to the possible existence of massless tadpoles in the theory.

Before we proceed let us point out that during the process of carrying out this
iterative procedure we shall encounter repeated operation of 
$(L_0^+)^{-1}(1-{\bf P})$
and $[\cdots]$, {\it e.g.} $[\cdots (L_0^+)^{-1}(1-{\bf P}) [\cdots]]$. 
Representing $1/L_0^+$ as $\int_0^\infty ds e^{-s L_0^+}$
and the $L_0^-=0$ projector as $\int d\theta e^{i\theta L_0^-}$ we can interpret
terms with repeated application of $1/L_0^+$ and $[~]$ in terms of correlation functions
on a single Riemann surface obtained by joining 1PI Riemann surfaces by the
plumbing fixture relations \refb{eplumb}. The $(1-{\bf P})$ factor plays a crucial
role: it subtracts off the contribution from the massless states in the propagator,
thereby removing possible divergences from the $s\to\infty$ end of the integral.
As a result this procedure does not require any infrared regulator.

Let us now turn to 
another consequence of the presence of non-trivial elements of BRST
cohomology in the $L_0^+=0$ sector. 
If
eq.\refb{eobstruct} holds then \refb{esol1} has solutions. 
Let 
$|\Phi^{(0)}_{k+1}\rangle$ be a particular solution 
which differs from $|\Phi_k\rangle$
by terms of order $g_s{}^{k+1}$.  However from this we can construct a
more general family of solutions to \refb{esol1} with the same property by taking
\be \label{ecni}
|\Phi_{k+1}\rangle = |\Phi_{k+1}^{(0)}\rangle + 
g_s{}^{k+1} \sum_i c^{(k+1)}_i |V_i\rangle\, ,
\ee
where $c^{(k+1)}_i$ are arbitrary constants.
This freedom exists at every order in $g_s$.
However since the choice of $c^{(\ell)}_i$'s for a given $\ell$ affects the solution
at higher order via the recursion relations,
it may so happen that \refb{eobstruct} fails to hold for arbitrary choice
of $c^{(\ell)}_i$'s at lower order. In that case we have to 
adjust 
the $ c^{(\ell)}_i$'s at lower order to make \refb{eobstruct} hold.
If this is not possible, i.e.\
we cannot make \refb{eobstruct} hold for any choice of $c^{(\ell)}_i$'s then
we have to conclude that 
the system does not have  a perturbative ground state. This is what happens if
there are tadpoles of classical moduli fields in the theory.
For theories in which
\refb{eobstruct} can be made to hold, we can
divide the massless states into two kinds. For some states
the constants $c^{(\ell)}_i$ remain undetermined at every order in $g_s$. These
represent scalar fields without potential -- the massless moduli fields. For 
other states the constants $c^{(\ell)}_i$ get fixed by requiring \refb{eobstruct} to hold.
These are scalar fields with potential.\footnote{Most often massless scalars which are
not moduli fields transform in the non-trivial representation of some gauge group and
hence we can get a solution by setting them to zero from the beginning. However,
as we shall discuss in \S\ref{sr.3}, there can some
time be more than one possible choices of the constants $c^{(\ell)}_i$, at least in
low orders in $g_s$, signalling
the existence of multiple perturbative vacuum. In this case the $c^{(\ell)}_i=0$ solution
may not represent the true ground state, i.e.\ \refb{eobstruct} at higher order may force us to
choose the $c^{(\ell)}_i\ne 0$ solution.}

Note that even if we ignore the freedom provided by the vacuum expectation values
of the moduli fields, the solution is ambiguous due the freedom
of adding BRST trivial state $Q_B|s_{k+1}\rangle$ in \refb{esoln}. 
Any such term added at a given order will of course affect
the higher order terms in the solution. It can be shown that this freedom reflects the
freedom of transforming a given solution by a gauge transformation of the form 
described in \refb{egauge}. Since this is gauge equivalent to the original solution,
we can set $|s_{k+1}\rangle$ to zero from the beginning.

Once we have obtained a classical solution -- which we shall denote by $\Psi_{cl}$
-- we can expand the action around the new background and obtain the new action.
After throwing away the additive constant $S(|\Psi_{cl}\rangle)$,
the new action can be expressed in terms of the fluctuation $\wh\Psi= \Psi - \Psi_{cl}$
as\cite{aseq}
\be \label{emodaction}
g_s{}^{-2}\left[ {1\over 2} \langle\wh\Psi|  c_0^- Q_B|\wh\Psi\rangle +
\sum_{n=2}^\infty {1\over n!} \, \{\wh\Psi^n\}'\right]\, ,
\ee
where we define
\ben \label{eredefined}
&& \{ A_1\cdots A_k\}' \equiv \sum_{n=0}^\infty {1\over n!} \, \{\Psi _{cl}^n A_1\cdots A_k\}\, ,
\qquad \hbox{for $k\ge 2$}\, ,
\nonumber \\
&& [A_1\cdots A_k]' \equiv \sum_{n=0}^\infty {1\over n!} \, [\Psi _{cl}^n A_1\cdots A_k]\, ,
\qquad \hbox{for $k\ge 1$}\, , \nonumber \\
&& \{A_1\}' \equiv 0, \qquad [~]'\equiv 0\, .
\een
Note the absence of linear term in $\wh\Psi$ in the action signalling that $|\wh\Psi\rangle=0$
is a solution of the equations of motion.
Furthermore one can show, using the equations of motion satisfied by
$| \Psi_{cl}\rangle$, that eqs.\refb{edefsquare}-\refb{emain}
hold with $\{\cdots\},[\cdots]$ replaced by $\{\cdots\}',[\cdots]'$. Thus the action
\refb{emodaction} has gauge invariance with $[\cdots]$ replaced by 
$[\cdots]'$ in the gauge transformation laws \refb{egauge}.

Due to the presence of the $\{\wh\Psi^2\}'$ term in the action \refb{emodaction}
the quadratic term
is not controlled only by $Q_B$. It is convenient to introduce a new operator $\wh Q_B$
defined via
\be \label{ewhQ}
\wh Q_B |A\rangle \equiv Q_B|A\rangle + [A]' 
= Q_B|A\rangle + \sum_{n=0}^\infty {1\over n!} \, [\Psi _{cl}^n A]\, .
\ee 
If we also define
\ben \label{eredefinedagain}
&& \{ A_1\cdots A_k\}'' \equiv \{ A_1\cdots A_k\}' \quad \hbox{for $k\ge 3$}, \qquad
[A_1\cdots A_k]''\equiv [A_1\cdots A_k]'  \quad \hbox{for $k\ge 2$}, \nonumber \\
&& \{A_1\}''\equiv 0, \quad \{A_1A_2\}''\equiv 0, \quad [~]''\equiv 0, \quad [A_1]''\equiv 0\, ,
\een
then one can express the action \refb{emodaction} as
\be \label{emodactionnew}
g_s{}^{-2}\left[ {1\over 2} \langle\wh\Psi|  c_0^- \wh Q_B|\wh\Psi\rangle +
\sum_{n=3}^\infty {1\over n!} \, \{\wh\Psi^n\}''\right]
\ee
so that there is no linear term and the quadratic term is controlled by $\wh Q_B$.
Furthermore one can show, using the equations of motion satisfied by
$| \Psi_{cl}\rangle$, that $\wh Q_B$ is nilpotent and that eqs.\refb{edefsquare}-\refb{emain}
hold with $Q_B,\{\cdots\},[\cdots]$ replaced by $\wh Q_B,\{\cdots\}'',[\cdots]''$. Thus the action
\refb{emodactionnew} has gauge invariance with $Q_B$ and $[\cdots]$ replaced by 
$\wh Q_B$
and $[\cdots]''$ in the gauge transformation laws \refb{egauge}. This generalizes the
result of \cite{aseq} for classical string field theory.

Before concluding this section we shall 
compare the procedure described above of shifting $\Psi$ by $\Psi_{cl}$
with the conventional 
approach to superstring perturbation theory where no such shift is needed.
As we have seen, the effect of shifting the background by $|\Psi_{cl}\rangle$
is to replace $\{~\}$ by $\{~\}'$. 
If in \refb{esoln} we had dropped the second and the third terms on the right hand side,
and also the projector $(1-{\bf P})$ from the first term, and used the corresponding
expression for $|\Psi_{cl}\rangle$ to define $\{~\}'$, then this would effectively
correspond to including
in the definition of
1PI amplitudes also 1PR contributions 
where the  internal propagators (i.e.\ punctures involved in the plumbing fixture)
carry zero momentum. By replacing $(L_0^+)^{-1}$ by
$\int_0^\infty ds\, e^{-s L_0^+}$ we can interpret these as the
`tadpole diagrams' of conventional perturbation theory.  
However in that case
one needs to use
suitable upper cut-off $\Lambda$ on the $s$ integral at the intermediate stage of the
calculation to tame the divergences coming from the $L_0^+=0$ 
states\cite{1209.5461}.
One then has to  check that at the end of
the calculation we can take the cut-off to infinity without encountering any divergence.
In the approach described above, where we use the full expression \refb{esoln},
we do not need any infrared regulator since effectively we subtract the
contribution from $L_0^+=0$ states from $\int_0^\infty ds\, e^{-s L_0^+}$, and then 
take into account the contribution from these missing states separately.
However in this approach we need to ensure that at each order
in $g_s$ we can find a solution to \refb{esol1} 
that can be used for constructing the solution
to the next order. This is equivalent to checking the  absence of tadpole divergence
in conventional perturbation theory.

\subsection{Effect of changing the local coordinates and/or PCO locations} \label{eeffct}

\begin{figure}

\begin{center}

\def\JPicScale{0.8}
\ifx\JPicScale\undefined\def\JPicScale{1}\fi
\unitlength \JPicScale mm
\begin{picture}(155,65)(0,0)
\linethickness{0.3mm}
\put(20,60){\line(1,0){120}}
\put(20,60){\vector(-1,0){0.12}}
\linethickness{0.3mm}
\put(20,40){\line(0,1){20}}
\put(20,40){\vector(0,-1){0.12}}
\linethickness{0.3mm}
\put(20,40){\line(1,0){120}}
\put(140,40){\vector(1,0){0.12}}
\linethickness{0.3mm}
\put(140,40){\line(0,1){20}}
\put(140,60){\vector(0,1){0.12}}
\put(80,65){\makebox(0,0)[cc]{$\RR'_{g,n}$}}

\put(80,50){\makebox(0,0)[cc]{$\wt\RR_{g,n}$}}

\put(80,35){\makebox(0,0)[cc]{$-\RR_{g,n}$}}

%\put(28,50){\makebox(0,0)[cc]{$(\p\RR)'$}}

%\put(131,50){\makebox(0,0)[cc]{$(\p\RR)'$}}

%\put(80,50){\makebox(0,0)[cc]{$\RR$}}

\linethickness{0.3mm}
\put(155,40){\line(0,1){20}}
\put(155,60){\vector(0,1){0.12}}
\put(156,35){\makebox(0,0)[cc]{$\wh U_{g,n}$}}

\end{picture}

\end{center}

\vskip -1.3in

\caption{A pictorial representation of eqs.\refb{edeltaactionpre} and \refb{edeltaaction}.
The right hand side of \refb{edeltaactionpre} which is also the left hand side
of \refb{edeltaaction} is the contribution to the boundary integral of
$\Omega^{(g,n)}_{6g-6+2n}$ from the upper and lower horizontal
edges of the rectangle.
The first term on the right hand side of \refb{edeltaaction} is the volume integral
of $d\Omega^{(g,n)}_{6g-6+2n}$ over the interior $\wt\RR_{g,n}$ of the rectangle. 
Since the height of the rectangle is infinitesimal we can replace the effect of
integration along the vertical direction by contraction with $\wh U_{g,n}$. 
Finally the last
term of \refb{edeltaaction} represents the negative of the contribution to the boundary integral
of $\Omega^{(g,n)}_{6g-6+2n}$ from the vertical edges of the rectangle. Thus 
\refb{edeltaaction} follows from Stoke's theorem. 
Although we have taken the height of the rectangle to be constant for the ease of drawing
the figure, this is certainly not necessary. Finally note that here we have drawn $\RR_{g,n}$
and $\RR'_{g,n}$ as one dimensional horizontal lines, but the general case corresponds to
them being multidimensional, with the whole figure stretching out of the plane of the 
paper / screen.
\label{fonly}}

\end{figure}

We now turn to the problem of studying the effect of changing the choice of
local coordinates and/or the locations of the PCO's on the 1PI action. A change of
this form will correspond to a new choice of the $(6g-6+2n)$ dimensional 
regions $\RR_{g,n}$ in $\wt\PP_{g,n}$
satisfying \refb{eboundary}. Let us denote them by $\RR'_{g,n}$. We shall consider
infinitesimal deformations so that $\RR_{g,n}$ and $\RR'_{g,n}$ are close in
$\wt\PP_{g,n}$. Then we can write
\be \label{edeltaactionpre}
\delta S = \sum_{g=0}^\infty  g_s{}^{2g-2}  \sum_{n=1}^\infty {1\over n!} \,
\left[\left(\int_{\RR'_{g,n}} - \int_{\RR_{g,n}}\right)
\, \Omega^{(g,n)}_{6g-6+2n}(|\Psi\rangle^{\otimes n})
\right]\, ,
\ee
where $|\Psi\rangle^{\otimes n}$ denotes that there are $n$ entries of
$|\Psi\rangle$ in the argument.
Let $\hat U_{g,n}$ be an infinitesimal vector field that takes a point in $\RR_{g,n}$ to a
neighbouring point in $\RR'_{g,n}$. The definition of $\hat U_{g,n}$ is ambiguous up to 
addition of infinitesimal tangent vectors of $\RR_{g,n}$, but this will not affect the
final result. 
In this case \refb{edeltaactionpre} can be expressed as\cite{9301097}
\be \label{edeltaaction}
\delta S = \sum_{g=0}^\infty  g_s{}^{2g-2}  \sum_{n=1}^\infty {1\over n!} \, 
\left[\int_{\RR_{g,n}}
\, d\Omega^{(g,n)}_{6g-6+2n}[\hat U_{g,n}] (|\Psi\rangle^{\otimes n}) \right.  %\nonumber \\ &&
 \left. +
\int_{\p \RR_{g,n}} \Omega^{(g,n)}_{6g-6+2n}[\hat U_{g,n}] (|\Psi\rangle^{\otimes n}) 
\right]\, ,
\ee
where for any $p$-form $\omega_p$, 
$\omega_p[\hat U]$ denotes the contraction of 
$\omega_p$ with the vector field $\hat U$:
\be
\omega_{i_1\cdots i_p}dy^{i_1} \wedge \cdots \wedge dy^{i_p}[\wh U] \equiv \wh U^{i_1} 
\omega_{i_1 i_2\cdots i_p}dy^{i_2} \wedge \cdots \wedge dy^{i_p}\, .
\ee
Intuitively this equation can be understood as follows. The first term on the right
hand side represents the integral of $d\Omega^{(g,n)}_{6g-6+2n}$ over a $6g-5+2n$
dimensional region $\wt \RR_{g,n}$
bounded by $\RR_{g,n}$ and $\RR'_{g,n}$. This can be integrated
to give \refb{edeltaactionpre} together with a contribution from the component of the
boundary of $\wt\RR_{g,n}$ that joins $\p\RR'_{g,n}$ to $\p\RR_{g,n}$. The second
term in \refb{edeltaaction} subtracts this contribution. A pictorial representation of this
can be found in Fig.~\ref{fonly}.

We shall now show following 
\cite{9301097} that the change in action given in \refb{edeltaaction}
can be regarded as the result of a redefinition of the field $|\Psi\rangle$ to
$|\Psi\rangle+|\tilde\delta \Psi\rangle$ where $|\tilde\delta\Psi\rangle$ is given by
\be \label{edeltafield}
\langle \Phi| c_0^- |\tilde\delta \Psi\rangle 
=- \sum_{g=0}^\infty g_s{}^{2g} \, \sum_{n=1}^\infty  {1\over (n-1)!} \, 
\int_{\RR_{g,n}} \, \Omega^{(g,n)}_{6g-5+2n}[\hat U_{g,n}](|\Phi\rangle, 
|\Psi\rangle^{\otimes (n-1)})\, ,
\ee
for any state $|\Phi\rangle$ in $\HH_0$.
We now see that since
this is integrated over $\RR_{g,n}$, adding a tangent vector of $\RR_{g,n}$ to
$\hat U_{g,n}$ will not change the integral.
To prove \refb{edeltafield}, let us
denote by $\tilde\delta S$ the change in the action induced by the field redefinition
\refb{edeltafield}. Then from \refb{eaction} we get
\be \label{etildeS}
\tilde\delta S = g_s{}^{-2} \left[\langle \Psi|c_0^- Q_B |\tilde\delta\Psi\rangle 
+ \sum_{n=1}^\infty {1\over (n-1)!}  \{\Psi^{n-1}\tilde\delta\Psi\}\right]\, .
\ee
The first term can be written as
\ben 
&& g_s^{-2}\langle Q_B\Psi| c_0^-|\tilde\delta \Psi \rangle =-
\sum_{g=0}^\infty g_s{}^{2g-2} \, \sum_{n=1}^\infty  {1\over (n-1)!} \, 
\int_{\RR_{g,n}} \, \Omega^{(g,n)}_{6g-5+2n}[\hat U_{g,n}](Q_B|\Psi\rangle, 
|\Psi\rangle^{\otimes (n-1)})\,  \nonumber \\
&=& \sum_{g=0}^\infty g_s{}^{2g-2} \, \sum_{n=1}^\infty  {1\over n!} \,
\int_{\RR_{g,n}} \, d\Omega^{(g,n)}_{6g-6+2n}[\hat U_{g,n}](|\Psi\rangle^{\otimes n})
\een
In going from the second to the final expression in the above equation we first
averaged over all the $n$ possible position of $Q_B|\Psi\rangle$ inside the
argument of $\Omega^{(g,n)}_{6g-5+2n}[\hat U_{g,n}]$, and then used
\refb{eomega}.
This agrees with the first term on the right hand side of \refb{edeltaaction}. Thus it
remains to show that the second term on the right hand side of \refb{etildeS} agrees
with the second term on the right hand side of \refb{edeltaaction}. 
Using \refb{eboundary} the latter can be
expressed as
\be \label{ee.1}
-{1\over 2} \sum_{g=0}^\infty  g_s{}^{2g-2}  \sum_{n=0}^\infty {1\over n!} \,
\sum_{g_1,g_2,n_1,n_2\atop g_1+g_2=g, n_1+n_2=n+2}
\int_{{\bf S} [\{\RR_{g_1,n_1} , \RR_{g_2,n_2}\}]}
\Omega^{(g,n)}_{6g-6+2n}[\hat U_{g,n}] (|\Psi\rangle^{\otimes n}) \, ,
\ee
up to an additive constant. The additive constant corresponds to the $n=0$ term in 
\refb{ee.1} and has no effect on any physical quantity.
We can now carry out the integral over the angular variable $\theta$ in the plumbing
fixture relation \refb{eplumb} and use the relation \refb{efactor}. On 
$\{\RR_{g_1,n_1}, \RR_{g_2,n_2}\}$ the infinitesimal vector field $\hat U_{g,n}$ can be written
as a sum of two vector fields -- one with support on $\RR_{g_1,n_1}$, characterizing the 
difference between $\RR'_{g_1.n_1}$ and $\RR_{g_1,n_1}$ and the other with
support on $\RR_{g_2,n_2}$, characterizing the 
difference between $\RR'_{g_2,n_2}$ and $\RR_{g_2,n_2}$. Now since \refb{ee.1}
is invariant under $(g_1,n_1)\leftrightarrow (g_2,n_2)$, the two terms give the same
contribution. Thus we can keep only one of the terms, {e.g.} where $\hat U_{g,n}$ has 
support on $\RR_{g_2,n_2}$ and multiply the result by a factor of 2. This allows
us to write \refb{ee.1} as
\ben \label{ee.2}
&& -\sum_{g=0}^\infty  g_s{}^{2g-2}  \sum_{n=0}^\infty {1\over n!} \, 
\sum_{g_1,g_2,n_1,n_2\atop g_1+g_2=g, n_1+n_2=n+2}
{\bf S}\left[ 
\int_{\RR_{g_1,n_1}}  \Omega^{(g_1,n_1)}_{6 g_1-6+2n_1}(|\Psi\rangle^{\otimes (n_1-1)}, |\vp_r\rangle) \right. \nonumber \\ && \left.
\int_{\RR_{g_2,n_2}}
\Omega^{(g_2,n_2)}_{6g_2-5+2n_2}[\hat U_{g_2,n_2}] (|\vp^r\rangle, 
|\Psi\rangle^{\otimes (n_2-1)}) \right]\, .
\een
In arriving at \refb{ee.2} we have taken into account the fact that besides the
sign given in \refb{efactor}, there is
an extra minus sign resulting from 
the fact that the operation of contracting a form with $\theta$ independent vector
field $\wh U$ with vanishing $\wh U^\theta$ and integrating the form
over $\theta$ anti-commute, {\it e.g.}
\be
\int_\theta (dy\wedge d\theta[\wh U]) = \wh U^y \int_\theta d\theta = 2\pi \wh U^y,
\quad (\int_\theta dy \wedge d\theta)[\wh U] = - 2\pi \, dy[\wh U] = -2\pi \wh U^y\, .
\ee
Now since all the external states are $|\Psi\rangle$, the ${\bf S}$ operation 
in \refb{ee.2} just
produces a factor of $n!/(n_1-1)! (n_2-1)!$. Regarding $(g_1,n_1)$ and $(g_2,n_2)$ as 
independent summation variables and performing the sum over $g_1$,
we can express \refb{ee.2} as
\ben \label{ee.3}
%&& 
-g_s{}^{-2} \, \sum_{n_1=1}^\infty {1\over (n_1-1)!} 
\{\Psi^{n_1-1} \vp_r\} %\nonumber \\ && \times 
\sum_{g_2=0}^\infty \sum_{n_2=1}^\infty {1\over (n_2-1)!}g_s^{2g_2} 
\int_{\RR_{g_2,n_2}}
\Omega^{(g_2,n_2)}_{6g_2-5+2n_2}[\hat U_{g_2,n_2}] (|\vp^r\rangle, 
|\Psi\rangle^{\otimes (n_2-1)})  \nonumber \\
%&=&  
=g_s{}^{-2} \sum_{n_1=1}^\infty {1\over (n_1-1)!} 
\{\Psi^{n_1-1} \vp_r\} \langle \vp^r|c_0^- | \tilde \delta \Psi\rangle
=  g_s{}^{-2} \sum_{n_1=1}^\infty {1\over (n_1-1)!}
\{\Psi^{n_1-1} \tilde \delta \Psi\}  \, . \qquad \qquad \qquad \quad %\nonumber \\
\een
In the last step we have used the relation $|\vp_r\rangle \langle \vp^r|c_0^-|s\rangle
= |s\rangle$ for $|s\rangle\in\HH_0$ -- this follows from \refb{ebasis}.
\refb{ee.3} agrees with the second term on the right hand side of \refb{etildeS},
establishing the equality of \refb{edeltaaction} and \refb{etildeS} up to an
additive constant.

Finally we note that even though we have described the field redefinition in terms of the
original string field $\Psi$, we can also translate this to a redefinition of the shifted
field $\wh\Psi$ using the known relation between $\wh\Psi$ and $\Psi$.

The result of this subsection and the previous one leads to the following question. 
Suppose that we have two different choices of $\RR_{g,n}$ leading to two different
actions. 
Suppose further that for each action 
we have constructed vacuum
solutions following the procedure described in \S\ref{esolution}.
If we now perform the field redefinition described in this subsection to relate
the two actions, does it map these
vacuum solutions to each other? If the vacuum solutions are unique up to gauge
transformations, then it is guaranteed that they will be mapped to each other
up to a gauge transformation. However if there are moduli fields whose vacuum 
expectation values are not fixed, then 
all that one can conclude is that the family of vacuum solutions for the two actions
will be related to each other under field redefinition, but the transformation rules for
the parameters $c^{(n)}_i$ in \refb{ecni} under this transformation will be very complicated
in general. In particular even if we take all the $c^{(n)}_i$'s to be zero while solving the equations 
of motion derived from the first action, 
it may not map to the solution with vanishing $c^{(n)}_i$'s in the family of
solutions to the equations of
motion derived from the second action.
This explains the observation of 
\cite{catoptric,1209.5461,1408.0571} that a change in the
local coordinate system and/or PCO locations will have to be accompanied by a 
shift in the expectation values of the moduli fields in order to ensure that we are in
the same physical theory, i.e.\ to keep the mass spectrum and 
S-matrix elements unchanged.

\sectiono{Applications} \label{sapp}

We shall now describe how the 1PI effective action can be used to simplify the
analysis of mass renormalization and vacuum shift in string theory. It should be
kept in mind however that since the 1PI effective action gives the same off-shell
amplitudes as those obtained using the prescription used in 
\cite{1311.1257,1401.7014,1404.6254,1408.0571}, the results
remain unchanged. Only the proof of some of the results, in particular the result
that physical quantities are independent of the choice of local coordinates and PCO
locations used to define the off-shell amplitudes, simplify.

\subsection{Mass renormalization} \label{sr.1}

We begin our discussion with the computation of the renormalized mass. 
Once we have found the vacuum, the 1PI action expanded around the vacuum has no
linear term. The quadratic terms define the kinetic operator $M$.
If we expand the string field in some basis $\{|\bar\vp_r\rangle\}$ of ghost number two
states in $\HH_0$, then the explicit form of $M_{rs}$ is given by
\be
M_{rs} = \langle\bar\vp_r| c_0^- \wh Q_B|\bar\vp_s\rangle \, ,
\ee
where $\wh Q_B$ has been defined in \refb{ewhQ}.
Now 
we examine the eigenvalues
of $M$.\footnote{Note that even though $\wh Q_B$ is nilpotent, the matrix 
$M$ is not nilpotent. Hence it can have non-zero eigenvalues.}
This can be done at fixed momentum along the non-compact
directions due to translational invariance along those directions, and then we can study how the
eigenvalues vary as
a function of the momentum. There are three kinds
of behaviour:
\begin{enumerate}
\item Some of the eigenvalues will vanish for all momenta. These are pure gauge
directions.
\item Some eigenvalues never vanish as a function of momentum. These are unphysical
modes.
\item Some eigenvalues vary as a function of momentum $k$ 
and vanish at specific values of
$k^2$. These describe physical states. The locations of the zeroes in the $-k^2$ plane
give the physical renormalized mass$^2$.
\end{enumerate}
As an example from field theory,
we can consider the free photon kinetic term in quantum electrodynamics.
In momentum space it is proportional to $(-k^2\eta^{\mu\nu} + k^\mu k^\nu)$. Taking
$\{k^\mu\}=(k^0,k^1,0,0)$ the kinetic operator takes the form
\be
\pmatrix{(k^1)^2  &  k^0 k^1 && \cr k^0 k^1 & (k^0)^2 && \cr && (k^0)^2- (k^1)^2 &\cr 
&&& (k^0)^2- (k^1)^2}\, .
\ee
This has eigenvalues
\be 
\lambda_1(k) = 0, \quad \lambda_2(k) = (k^0)^2+ (k^1)^2, \quad \lambda_3(k)
=\lambda_4(k)=(k^0)^2 - (k^1)^2\, .
\ee
Since $\lambda_1(k)$ vanishes for all $k$, it represents a pure gauge mode. 
$\lambda_2(k)$ never vanishes (except at the special point $k^0=k^1=0$) and hence
describes an unphysical mode. $\lambda_3(k)$ and $\lambda_4(k)$ vanish
at $k^0=k^1$, and describe massless physical states.
The important point to note is that the mass spectrum can be found without any 
gauge fixing.
If instead we had added a gauge fixing term and then computed the eigenvalues
of
the kinetic term and the locations of their zeroes, the result will in general depend on the
choice of gauge. We'll then have to make special effort to determine which of these
zeroes describe physical states (for which the location of the zeroes in the $k^2$ 
plane should not depend on the gauge choice) and which are gauge artifacts. This
was the main problem encountered in the analysis of \cite{1401.7014}. 
Here we see that the use
of gauge invariant 1PI effective action allows us to
circumvent this problem by working with the gauge invariant quantum
corrected kinetic term.

Now consider the effect of a change in the choice of local coordinates and/or
PCO locations. As discussed in \S\ref{eeffct}, this can be compensated by a field
redefinition under which the original vacuum gets mapped to a vacuum of the
new action for some specific choice of the moduli fields.
As a result the kinetic operator $M$ around the corresponding vacua are related to each other
by a transformation of the form
\be 
M \to A(-k)^T M A(k)\, ,
\ee
where $A(k)$ is an operator acting on states of ghost number two and momentum $k$ in
$\HH_0$ that can be computed using the 
field redefinition \refb{edeltafield}. The important point to note is that since
\refb{edeltafield} involves integration over the 1PI regions $\RR_{g,n}$, it does
not have any pole in the $-k^2$ plane in perturbation theory. Thus for every 
eigenstate
of zero eigenvalue of $M$, we can construct an eigenstate of zero eigenvalue of
$A(-k)^TMA(k)$ by multiplying the original eigenstate by the non-singular matrix
$A(k)^{-1}$. This shows that the
locations of the zeroes of the eigenvalues in the $-k^2$ plane are not affected by
the field redefinition. This is turn establishes that the physical renormalized masses
are independent of the choice of local coordinate system and the locations of the
PCO's.  

The above discussion has been somewhat formal, in the sense that it requires
us to work with infinite dimensional matrices $M$. We shall now describe
how to reduce the problem of computing renormalized mass 
to a more manageable form by `integrating out' contribution from 
all states except those at a given mass level $m$ -- where mass level of a state 
carrying momentum $k$ is defined by the condition that its $L_0^+$ eigenvalue vanishes
for $k^2=-m^2$. This will then allow us to work with states of a given mass level
at a time, which are finite in number. During this analysis we shall also 
develop a systematic perturbation expansion for computing the renormalized
masses.

We recall that in order to address the problem of mass
renormalization we have to find zero eigenvalues of $M$, i.e. find
solutions to the equation 
\be \label{eqexp}
\wh Q_B |\psi\rangle=0\, .
\ee
Let us express $\wh Q_B$ as
\be
\wh Q_B = Q_B + K, \qquad K |A\rangle \equiv \sum_{n=0}^\infty {1\over n!} \, [\Psi_{cl}^n A]
\ee
so that $K$ contains operators of order $g_s{}^2$ and higher. 
Since the natural expansion parameter is $\kappa\equiv g_s{}^2$, 
we shall from now on
express all the quantities as a power series expansion in $\kappa$.\footnote{Here we
are ignoring the special cases discussed in \S\ref{sr.3} where $|\Psi_{cl}\rangle$ can be
of order $g_s$ leading to $g_s$ as the expansion parameter. Our analysis can be
extended to these  cases as well.}
We shall denote by $|\psi_n\rangle$ the result for $|\psi\rangle$ accurate to order 
$\kappa^n$, and the same convention will be followed for all other states. 
Then we can express \refb{eqexp} as
\be \label{esa1}
Q_B|\psi_{n+1}\rangle = - K\, |\psi_n\rangle + \OO(\kappa{}^{n+2})\, .
\ee
Let us suppose that we have found $|\psi_n\rangle$ satisfying \refb{esa1} with
$n$ replaced by $(n-1)$. Then
using the nilpotence of $Q_B+K$ one can show that 
$Q_B K|\psi_n\rangle=\OO\left(\kappa{}^{n+2}\right)$. We can now 
write down a formal solution
to \refb{esa1} of the form\footnote{A more general solution to \refb{esa1} will
allow us to add a term of the form $Q_B|\chi_{n+1}\rangle$
to the right hand side of \refb{esa3}.
We are allowed to drop such terms this since we are interested in
finding solutions to \refb{esa1} which are not pure gauge deformations.}
\be \label{esa2}
|\psi_{n+1}\rangle = -{b_0^+\over L_0^+} K |\psi_n\rangle + \OO\left(\kappa{}^{n+2}\right)\, .
\ee
\refb{esa2} makes sense as long as $K|\psi_n\rangle$ is a linear combination of states
with $L_0^+\ne 0$.
Now while applying the above procedure we shall always begin with an initial state $|\psi_0\rangle$
at some given mass level $m$ and proceed. On shell condition at tree level then requires the
momentum carried by
the state to satisfy $k^2=-m^2$ i.e.\ $L_0^+=0$. 
Since in perturbation theory we shall
keep the momentum
$k$ close to the original value we see that for states at the same mass level $m$ as 
$|\psi_0\rangle$, $L_0^+$ eigenvalues will be small (of order $\kappa$) 
and hence the operator ${b_0^+\over L_0^+} K$ may be of order one, signalling a breakdown
of the perturbative procedure. For this purpose we shall introduce a projection operator 
$P$ that projects onto states of mass level $m$ and apply this recursive technique
only on states other than those at mass level $m$. Let us suppose that we want
to compute the renormalized masses accurately up to order $\kappa^N$.
Then we claim that the following is a solution to \refb{esa1} for $n\le N-1$
\be \label{esa3}
|\psi_0\rangle = |\phi_N\rangle, \qquad
|\psi_{n+1}\rangle = -{b_0^+\over L_0^+} (1-P) K |\psi_n\rangle + |\phi_{N}\rangle + 
\OO\left(\kappa{}^{n+2}\right)\, , 
\ee
where $|\phi_{N}\rangle$ satisfies
\be \label{esa4}
P|\phi_{N}\rangle = |\phi_{N}\rangle\, , 
\ee
\be \label{esa5}
Q_B|\phi_{N}\rangle = - P 
K |\psi_{N-1}\rangle +\OO(\kappa^{N+1}) \, .
\ee
The proof that \refb{esa3} satisfies \refb{esa1} goes as follows.
The projection condition \refb{esa4} tells us that $|\phi_{N}\rangle$ is  a 
level $m$ state.\footnote{Notice 
that we have not introduced the states $|\phi_n\rangle$ for $0\le n\le N-1$ which
would provide approximations to $|\phi_N\rangle$ to order $\kappa^n$. Instead of
determining $|\phi_n\rangle$ perturbatively, we shall later determine it in a single 
step.} Using \refb{esa3} to express $|\psi_{n+1}\rangle - |\psi_n\rangle$ in terms of
$|\psi_n\rangle -|\psi_{n-1}\rangle$ and noting that $|\psi_1\rangle -\psi_0\rangle$
is of order $\kappa$ one can show iteratively that
$|\psi_{\ell+1}\rangle -|\psi_\ell\rangle
\sim \kappa^{\ell+1}$ for all $\ell$. This observation, together with eq.\refb{esa5},
gives $Q_B|\phi_{N}\rangle = - P 
K |\psi_{n}\rangle +\OO(\kappa^{n+2})$. Using this and assuming that \refb{esa1} holds
with $n$ replaced by $(n-1)$ one can easily verify that \refb{esa3} satisfies
\refb{esa1}. 

By iterating this solution till $n=N-1$ starting with the seed solution 
$|\psi_0\rangle=|\phi_N\rangle$
we can determine $|\psi_n\rangle$ in terms of $|\phi_N\rangle$ for $0\le n\le N$.
In particular we get an expression for $|\psi_{N-1}\rangle$ in terms
of $|\phi_N\rangle$ of the form
\be 
|\psi_{N-1}\rangle = S|\phi_N\rangle + \OO(\kappa^N)
\ee
for some linear operator $S$.
$S$ involves multiple successive operations of $(1-P)(L_0^+)^{-1}$
and $K$, and, as in \S\ref{esolution}, each term involving such products of operators
may be interpreted as contribution from a
single Riemann surface obtained by joining 1PI Riemann surfaces 
via plumbing fixture. The
effect of the $(1-P)$ operator is to remove the contribution from the states of mass
level $m$ from the propagator which could generate large numbers of order $1/\kappa$
from the $s$ integral.

Eq.\refb{esa5} now gives
\be \label{efineq}
Q_B|\phi_{N}\rangle = - P K S |\phi_{N}\rangle+\OO(\kappa^{N+1}) \, .
\ee
Since $|\phi_N\rangle$ is an arbitrary state of mass level $m$, we can express this
as a linear combination of the basis states $|r\rangle\in\HH_0$ of ghost number two
and  mass level $m$:
\be
|\phi_N\rangle = \sum_r v_r |r\rangle\, .
\ee 
Taking
the inner product of \refb{efineq} with $\langle s| c_0^-$ where $\langle s|$
is the BPZ conjugate of $|s\rangle$,  we get 
\be
\sum_r \langle s| c_0^- (Q_B + P K S) |r\rangle \, v_r = \OO(\kappa^{N+1})\, .
\ee
Thus the problem reduces to finding the zero eigenvalues of a finite dimensional 
matrix.
Solutions which exist for all $k$ will represent pure
gauge states while solutions which exist only for some fixed value of $k^2$
near $-m^2$ will represent physical states.
Explicit form of pure gauge solutions will be given in \refb{esa7}, \refb{esa8}.

In the next subsection we shall describe the relationship of this approach to the
Siegel gauge analysis of \cite{1401.7014}.

\subsection{Relation to Siegel gauge analysis} \label{sr.1.5}

Ref.\cite{1401.7014} gave a 
prescription for computing mass renormalization,
but it was based on the analysis of Siegel gauge propagator and hence is
apparently different from the definition of physical renormalized mass we have 
suggested
above. We shall now show that the definition given in \S\ref{sr.1} agrees with the one in
\cite{1401.7014}, and hence we can use the argument given in \S\ref{sr.1}
to conclude
that the perturbative method described in \cite{1401.7014} also gives results which
are independent of the choice of local coordinate system and locations of PCO's.

Our strategy will be to show that any solution to \refb{esa1} can be made to satisfy the
Siegel gauge condition after adding to it a pure gauge state.
This will justify the approach of \cite{1401.7014} which worked with states in the Siegel gauge.
First we note that by construction $|\psi_{n+1}\rangle$ given in \refb{esa3} 
satisfies the Siegel gauge
condition for states at level other than $m$, 
i.e.\ satisfies
\be \label{esa6}
b_0^+ (1-P) |\psi_{n+1}\rangle = \OO(\kappa^{n+2})\, ,
\ee
Thus we need to show that $|\phi_{N}\rangle$ can also be made to satisfy
Siegel gauge condition.
We shall do this by exploiting 
the freedom of adding
pure gauge solutions to $|\psi_{n+1}\rangle$, but for this we must first
find the form of the pure gauge solutions satisfying \refb{esa6}.
We shall now describe an algorithm to generate such solutions.
We first construct $|\lambda_n\rangle$ for $0\le n\le N$ by
solving the recursion relations
\be \label{esa7}
|\lambda_0\rangle =|\eta_N\rangle, \quad
|\lambda_{n+1}\rangle = - {b_0^+\over L_0^+} (1-P) K |\lambda_n\rangle + |\eta_{N}\rangle
+ \OO\left( \kappa{}^{n+2}\right)\, ,
\ee
where $|\eta_{N}\rangle$ is an arbitrary state of ghost number 1, 
momentum $k$
and mass level $m$ in $\HH_0$, with its $\kappa$ expansion beginning at
$\kappa^0$ and going up to $\kappa^{N}$.  Using \refb{esa7} to express
$|\lambda_{n+1}\rangle -|\lambda_n\rangle$ in terms of $|\lambda_n\rangle
-|\lambda_{n-1}\rangle$ and noting that $|\lambda_1\rangle$ and $|\lambda_0\rangle$
differ by order $\kappa$, one can show iteratively that 
$|\lambda_{n+1}\rangle -|\lambda_n\rangle=\OO(\kappa^{n+1})$.
We now
define
\be \label{esa8}
|\psi^g_{n+1}\rangle = (Q_B+K) |\lambda_{n+1}\rangle + \OO(\kappa{}^{n+2})
= Q_B |\lambda_{n+1}\rangle + K |\lambda_{n}\rangle+ \OO(\kappa{}^{n+2})
\quad \hbox{for $0\le n+1 \le N$}
\, .
\ee
It follows using the nilpotence of $(Q_B+K)$ that $|\psi^g_{n+1}\rangle$ satisfies 
\refb{esa1} at generic momentum and hence is a pure gauge state. 
Furthermore from \refb{esa7} it is easy to check that $|\psi^g_{n+1}\rangle$
satisfies \refb{esa6}.

Let us now focus our attention 
on the physical states. Suppose 
that $m_R=m+\OO(\kappa)$ is the correct renormalized mass for some 
physical state. Then at $k^2=-m_R{}^2$ 
it will have the general form given in \refb{esa3}-\refb{esa5}. 
Now if we add to it any  pure gauge state of the from 
\refb{esa8} carrying the same momentum, it will continue to remain an
eigenstate of the kinetic operator 
with zero eigenvalue.  
Our goal will be to argue that by adding appropriate pure gauge states of the form
\refb{esa8} we can make each of the renormalized physical states satisfy Siegel gauge 
condition.
Since we shall be computing the solution
accurate up to order $\kappa^N$ we shall take a pure gauge solution up to that order.
Thus the general solution takes the form
\be \label{ephys}
|\chi_{N}\rangle \equiv |\psi_{N}\rangle + (Q_B+K)|\lambda_{N}\rangle
= |\psi_{N}\rangle + Q_B|\lambda_{N}\rangle +K |\lambda_{N-1}\rangle
+ \OO(\kappa^{N+1})
\ee
where $|\psi_{n+1}\rangle$ and $|\lambda_{n+1}\rangle$ satisfy the
recursion relations \refb{esa3} and \refb{esa7} for $0\le n\le (N-1)$:
\ben \label{erecur}
&& |\psi_0\rangle = |\phi_N\rangle, \qquad |\psi_{n+1}\rangle 
= -{b_0^+\over L_0^+} (1-P) K |\psi_n\rangle + |\phi_{N}\rangle + 
\OO\left(\kappa{}^{n+2}\right), \nonumber \\ &&
|\lambda_0\rangle = |\eta_N\rangle, \qquad 
|\lambda_{n+1}\rangle = - {b_0^+\over L_0^+} (1-P) K |\lambda_n\rangle + |\eta_{N}\rangle
+ \OO\left( \kappa{}^{n+2}\right)\, .
\een
Using \refb{ephys}, \refb{erecur} we now get
\be \label{esiegel}
b_0^+|\chi_{N}\rangle = b_0^+(|\phi_{N}\rangle +  P \, K\, |\lambda_{N-1}\rangle 
+ Q_B \, |\eta_{N}\rangle)+\OO(\kappa^{N+1})\, .
\ee
Now it follows from \refb{esa5} that 
$Q_B|\phi_{N}\rangle$ has its expansion starting at order $\kappa$.
Thus in the $\kappa\to 0$ limit $|\phi_{N}\rangle$
is a BRST invariant state.
It then follows from the general result on BRST cohomology that
\be
|\phi_{N}\rangle 
= c\bar c e^{-\phi} V (0)|0\rangle + Q_B|s\rangle + \kappa |\xi_{N-1}\rangle\, .
\ee
Here $V$ is a $\kappa$ independent 
matter vertex operator of dimension 
$(1/2+\alpha'(k^2+m^2)/4,1+\alpha'(k^2+m^2)/4)$
carrying momentum $k$, and becomes a superconformal primary operator in the
$k^2\to -m^2$ limit. 
$|s\rangle$ is some $\kappa$ independent state of level $m$ and
$|\xi_{N-1}\rangle$ is  a state of level $m$ whose $\kappa$ expansion starts
at order $\kappa^0$. Now by choosing
\be \label{elambda0}
|\eta_{N}\rangle = -|s\rangle - {b_0^+\over L_0^+} \left(\kappa |\xi_{N-1}\rangle
+ PK|\lambda_{N-1}\rangle\right)
\ee
we can make the right hand side of \refb{esiegel} vanish to order $\kappa^N$, 
showing that 
$|\chi_{N}\rangle$ satisfies the Siegel gauge condition. 
Since the  
$|\lambda_{N-1}\rangle$ determined from the recursion relation 
\refb{erecur} is a function of $|\eta_{N}\rangle$,
\refb{elambda0} gives a linear equation for $|\eta_{N}\rangle$ which needs
to be solved. 
Furthermore this has to be done by regarding this as a matrix equation in the
space of mass level $m$ states and not perturbatively,
since acting on
states of mass level $m$, $b_0^+ (L_0^+)^{-1}$ is of order $1/\kappa$ and hence
each of the terms in \refb{elambda0}
will have their $\kappa$ expansion beginning at order $\kappa^0$.
For example for $N=1$ we have $|\lambda_0\rangle=|\eta_1\rangle$
and hence \refb{elambda0} takes the form
\be
|\eta_1\rangle = -|s\rangle - {b_0^+\over L_0^+} \left(\kappa |\xi_{0}\rangle
+ PK|\eta_1\rangle\right)\, ,
\ee
i.e.
\be
|\eta_1\rangle = -\left( 1+ {b_0^+\over L_0^+} PK\right)^{-1} 
\left(|s\rangle + \kappa \,  {b_0^+\over L_0^+} |\xi_0\rangle\right)\, .
\ee
Furthermore using \refb{ephys}, \refb{erecur} 
we now get the leading order 
contribution to $|\chi_N\rangle$ to be
\be \label{efin}
|\chi_{N}\rangle = |\phi_{N}\rangle + Q_B|\eta_{N}\rangle +\OO(\kappa)
= c\bar c e^{-\phi} V(0) |0\rangle - Q_B
{b_0^+\over L_0^+}  (PK |\eta_N\rangle + \kappa |\xi_{N-1}\rangle)
+\OO(\kappa) \, .
\ee

This analysis shows that by adding appropriate pure gauge states we can
ensure that the renormalized physical state satisfies the Siegel gauge condition at
$k^2=-m_R{}^2$. This in turn implies that we can, from the beginning,
look for the physical states by working with states satisfying Siegel gauge
condition. However this is still not equivalent to
working with the Siegel gauge fixed 
action and studying its (linearized) equations of motion. In the latter case
we not only restrict the states to be
in the Siegel gauge, we also only examine 
a subset of the equations of motion -- namely those which have components
along the states in the Siegel gauge. Thus finding a zero eigenvalue of 
the Siegel gauge kinetic operator 
is a necessary but not sufficient condition for a physical state, and we need to find
some additional criterion that will tell us which of the zero eigenvalues of the
Siegel gauge kinetic operator describe
physical states.
Eq.\refb{efin} tells us how to identify these physical states in
the Siegel gauge -- in the 
$\kappa\to 0$ limit they must be given by a sum of a state of the form
$c\bar c e^{-\phi} V (0)|0\rangle$ and a BRST trivial state. In the Siegel gauge
analysis of \cite{1404.6254} it was found that if at tree level there are $n_p$ physical
states at mass level $m$, then after taking into account loop effects there are
precisely $n_p$ states with zero eigenvalue of the kinetic operator in the Siegel
gauge whose eigenstates have the form given in \refb{efin}. 
There were also other zero eigenvalues whose eigenstates were not of this form.
Thus the former must be 
the desired physical 
states, while the latter must be states which will fail to satisfy the linearized
equations of motion when we examine its components along states outside the
Siegel gauge. Indeed \cite{1404.6254} also identified the former 
as the true physical states using a
different set of arguments. The current analysis shows that these are true
eigenstates with zero eigenvalue 
of the full kinetic operator $c_0^-\wh Q_B$ and hence the renormalized
masses associated with these eigenstates are invariant under a change of local
coordinate system and / or PCO locations.

\subsection{S-matrix} \label{sr.2}

In order to compute the S-matrix from the 1PI action one needs to fix a gauge,
compute the gauge fixed propagator, and then compute the {\it tree level}
S-matrix in the theory using standard rules. While this does require a gauge
fixing and hence introduces possible dependence on spurious data in the intermediate
stage, we do not need to introduce any ghosts associated with the gauge fields
since we only compute tree amplitudes. Thus the standard proof that the tree level
S-matrix of a field theory is independent of the choice of gauge and field 
redefinition should hold.
The difficult part of the proof of gauge invariance (and invariance under field redefinition)
of the S-matrix elements, involving analysis of the invariance of the path integral measure
under these transformations, can be avoided altogether by working with gauge invariant 
1PI effective action.

\subsection{Vacuum shift} \label{sr.3}

We shall now turn to the analysis of vacuum shift studied in
\cite{1401.7014}. The set up is as follows. Suppose we have a scalar field $\phi$ that
is massless at the tree level but has a tree level potential $A\, g_s{}^{-2}\phi^4$ for some positive
constant
$A$. Suppose further that one loop correction generates a negative contribution
$-B\phi^2$ to the potential for some positive constant $B$. 
Thus we have the full potential 
\be
A\, g_s{}^{-2} \phi^4 - B\, \phi^2 + \cdots 
\ee
where $\cdots$ denotes other terms involving higher powers of $\phi$ and/or
of $g_s{}^2$. Now it is clear that the potential has a minimum at 
\be
\phi^2 = Bg_s^2/2A + \cdots \, .
\ee
The question is: how do we study perturbative string theory around this minimum?

From the point of view of the 1PI effective action the answer is as follows.
Since $\phi$ is massless at the tree level, 
one of the $c_i$'s (or more generally some combination of the $c_i$'s)
appearing in \refb{ecni}
represents the vacuum expectation value of $\phi$.
While
solving the equations of motion to find $|\Psi_{cl}\rangle$, we should run into a situation
where the vanishing of \refb{eobstruct} can be achieved for three
possible solutions corresponding to $\phi=0$ and 
$\phi=\pm g_s\sqrt{B/2A}$.\footnote{Note that in general $\phi$ and $c_i$'s 
will be related by complicated functions, but to order $g_s$ we can expect the
relationship to be linear.}
Here we need to pick the solution for $c_i$'s that
corresponds to the minimum of the potential at $\phi=\pm g_s\sqrt{B/2A}$
rather than the maximum at $\phi=0$. Beginning with this solution we
need to systematically compute the higher order terms and correct the
solution at each order to ensure the vanishing of \refb{eobstruct}.
The actual procedure is similar to the one adopted in \cite{mukherji}, although the latter
paper only examined solutions to the tree level string field theory action.

Once we have found the solution to any given order we can
expand the action around the minimum as in \S\ref{esolution} and 
carry out the usual computation of renormalized masses and S-matrix elements.
The result will formally
be identical to the ones described in \cite{1404.6254}; however unlike in
\cite{1404.6254} here we shall not need to regulate the infrared divergence associated with
tadpoles of $\phi$ even at the intermediate stage, since we determine the vacuum
expectation value of $\phi$ (and other fields)
by solving the classical equations of motion rather than demanding the vanishing
of the $\phi$-tadpole. The invariance of the renormalized mass and S-matrix elements
under a change of local coordinate system and/or PCO locations now follow in
the same way as in \S\ref{sr.1}, \S\ref{sr.2}.

\bigskip

{\bf Acknowledgement:}
We thank Nathan Berkovits, Roji Pius, Arnab Rudra, Edward Witten
and Barton Zwiebach for useful discussions and Barton Zwiebach for his
valuable comments on an earlier version of the manuscript.
This work  was
supported in part by the 
DAE project 12-R\&D-HRI-5.02-0303 and J. C. Bose fellowship of 
the Department of Science and Technology, India.

\appendix

\sectiono{Issues with vertical segment} \label{svert}

In the analysis of \cite{1408.0571}
it was found that $\RR_{g,n}$'s cannot be taken to be sections over $\KK_{g,n}$
in the strict sense --
there are codimension one subspaces $\wt\KK_{g,n}$ of $\KK_{g,n}$ on which the locations of
the PCO's will have to change from an initial configuration $(z_1,\cdots z_p)$
to a final configuration $(z_1',\cdots z'_p)$ keeping the moduli fixed. 
In other words $\RR_{g,n}$ contains {\it vertical segments}.
Even though
the path connecting $(z_1,\cdots z_p)$
to  $(z_1',\cdots z'_p)$ passes through
the spurious singularities, there is a well defined value of the integral of
$\Omega^{(g,n)}_p$ along this path provided {\it the PCO locations 
are moved one at a time}.
Now we can consider two such paths in which the order in which the PCO locations
are moved differ. Each of these paths describe a choice of $\RR_{g,n}$. Let us call them
$\RR_{g,n}$ and $\RR'_{g,n}$. 
We would like to show that the actions corresponding to the choices $\RR_{g,n}$ and
$\RR'_{g,n}$ are related by a field redefinition. However the problem is that the change
is not infinitesimal and hence we cannot directly apply the analysis of 
\S\ref{eeffct}. We shall now discuss a way to circumvent this problem.

We 
recall that we are allowed to choose weighted averages of different PCO prescriptions
in defining the region $\RR_{g,n}$.
Thus we could take a one parameter family of regions $\RR_{g,n}(t)$ of the
form
\be 
\RR_{g,n}(t) = \sum_i f_{g,n}^{(i)}(t) \, \RR^{(i)}_{g,n}\, ,
\ee
where $\RR^{(i)}_{g,n}$ are different choices of $\RR_{g,n}$ differing in their 
vertical segments and 
$f_{g,n}^{(i)}(t) $'s are appropriate weight factors labelled by a parameter
$t$ satisfying 
\be \label{esum}
\sum_i f_{g,n}^{(i)}(t)=1
\ee 
and the gluing compatibility condition
\be \label{egcp}
\p \RR_{g,n}(t) = -{1\over 2} 
\sum_{g_1,g_1,n_1,n_2\atop g_1+g_2=g, n_1+n_2=n+2} 
{\bf S}\left[ \{\RR_{g_1,n_1}(t), \RR_{g_2,n_2}(t)\}
\right]
\ee 
for each $t$. We choose the $f_{g,n}^{(i)}(t)$ such that at $t=0$ and $1$ we
get back $\RR_{g,n}$ and $\RR'_{g,n}$ respectively, so that as we
deform $t$ from 0 to 1 we interpolate between the two choices of the vertical
segment which we wanted to show are equivalent. 
Now, under an
infinitesimal change $\delta t$ of $t$, we get
\be \label{eright}
\delta \RR_{g,n}(t) = \delta t \sum_i f_{g,n}^{(i)}{}'(t) \, \RR^{(i)}_{g,n}\, ,
\ee
where $'$ denotes derivative with respect to $t$.
The right hand side of \refb{eright} is a weighted sum of 
subspaces of $\wt\PP_{g,n}$.  Since $\delta\RR_{g,n}$ is infinitesimal, we might
hope to apply techniques similar to those in \S\ref{eeffct} to show that the change
in the action can be absorbed by a field redefinition. Our goal will be to show that this
can indeed be done. Once we establish this for infinitesimal deformations, the results
for finite deformation will follow.

Note that we cannot take
the interpolating region to be simply $(1-t)\, \RR^{(1)}_{g,n}+t\, \RR^{(2)}_{g,n}$ as this will
violate the gluing compatibility condition \refb{eboundary}.
To illustrate this let us suppose that the vertical segment first appears at genus
$g_0$ with $n_0$ punctures in the interior of $\KK_{g_0,n_0}$. If $\RR_{g_0,n_0}$
and $\RR'_{g_0,n_0}$ denote two such choices then we can take the
interpolating integration cycles to be $\RR_{g_0,n_0}(t)=
t\RR_{g_0,n_0}+(1-t)\RR_{g_0,n_0}$. But now we have
\be
\{\RR_{g_0,n_0}(t), \RR_{g_0.n_0}(t)\}
= t^2 \{\RR_{g_0,n_0}, \RR_{g_0.n_0}\} + 2 t (1-t) \{\RR_{g_0,n_0}, \RR'_{g_0.n_0}\}
+ (1-t)^2 \{\RR'_{g_0,n_0}, \RR'_{g_0.n_0}\}\, .
\ee
Since this forms a boundary of $\RR_{2g_0, 2n_0-2}$ we see that the choice of the
interpolating integration cycle  $\RR_{2g_0, 2n_0-2}(t)$ will have to 
be more complicated.
Similarly $\RR_{3g_0,3n_0-4}(t)$, which contains 
$\{\RR_{g_0, n_0}(t), \RR_{2g_0,2g_0}(t)\}$ as a boundary, must  be even more
complicated.

We now introduce the following (formal weighted sum of) subspaces of $\wt\PP_{g,n}$:
\begin{enumerate}
\item $\delta \RR_{g,n}(t)$ defined in \refb{eright} represents a formal sum of subspaces
of $\wt\PP_{g,n}$.
Due to \refb{esum} we have $\sum_i f_{g,n}^{(i)}{}'(t) =0$. Since $\RR^{(i)}_{g,n}$'s
differ from each other only on the fiber over $\wt\KK_{g,n}$, 
we see that $\delta \RR_{g,n}(t)$
lies inside the fiber over $\wt\KK_{g,n}$. Furthermore
since the restriction of
$\RR^{(i)}_{g,n}$ to the fiber over any point $m$ in $\wt\KK_{g,n}$ describe a path
in the fiber with the same initial and final points for each $i$, 
the 
$\sum_i f_{g,n}^{(i)}{}'(t) =0$ relation implies that 
the restriction of
$\delta \RR_{g,n}(t)$ to this fiber will be
a formal sum of closed paths in the fiber, or more specifically 
in the space of PCO locations at fixed choices of the local coordinate 
systems.\footnote{At the intersection of two such $\wt\KK_{g,n}$'s we have vertical
segments which are two dimensional surfaces in the space of PCO locations
instead of paths; on higher codimension surfaces involving intersection of multiple
$\wt\KK_{g,n}$'s the vertical segment has even higher dimensions. There are additional
subtleties involving these subspaces but they can be taken care of\cite{unpub}.}
We shall denote this formal sum of closed curves
by $\CC_{g,n}(m)$ for every $m\in \wt\KK_{g,n}$.  $\delta \RR_{g,n}(t)$
may now be identified as the collection of $\CC_{g,n}(m)$'s for all $m\in \wt\KK_{g,n}$.
\item $\CC_{g,n}(m)$ in turn may be regarded as the boundary of $\DD_{g,n}(m)$ --
a formal sum of two dimensional subspaces in the
fiber over $m$. We can take this two dimensional subspace  to lie along fixed
choices of local coordinate system, letting the PCO locations vary along $\DD_{g,n}(m)$.
\item We define
$\Delta_{g,n}$ to be the formal sum of $6g-5+2n$ dimensional subspaces of
$\wt\PP_{g,n}$ obtained by taking the collection of the $\DD_{g,n}(m)$'s
for all $m\in\wt\KK_{g,n}$.
\item We also define $\Gamma_{g,n}$ to be the formal sum of
$6g-6+2n$ dimensional subspace of
$\wt\PP_{g,n}$ obtained by taking the collection of the $\DD_{g,n}(m)$'s
for all $m\in \p\wt\KK_{g,n}$. Here $\p\wt\KK_{g,n}$ is the boundary of
$\wt\KK_{g,n}$ given by the intersection of $\wt\KK_{g,n}$ and
$\p\KK_{g,n}$.
\end{enumerate}
It follows from the above definitions that
the boundary of $\Delta_{g,n}$ is given by
\be \label{epdgn}
\p\Delta_{g,n} = \delta \RR_{g,n}(t)  +
\Gamma_{g,n}\, .
\ee 
Physically this reflects the fact that $\Delta_{g,n}$ has two kinds of boundaries.
If we consider a fixed point $m$ in $\wt\KK_{g,n}$ and move along the fiber
direction, we encounter the boundary $\CC_{g,n}(m)$, whose collection for all
$m\in\wt\KK$ gives the first term on the right hand side of \refb{epdgn}. On the
other hand if we move along $\wt\KK_{g,n}$ then we may encounter the boundary
$\p\wt\KK_{g,n}$. Thus the collection of $\DD_{g,n}(m)$ for $m\in \p\wt\KK_{g,n}$
gives the second boundary of $\Delta_{g,n}$ represented by the $\Gamma_{g,n}$ term
in \refb{epdgn}. 

As a consequence of the gluing compatibility relation \refb{egcp}
one can show that
\be
\Gamma_{g,n} = - \sum_{g_1,g_1,n_1,n_2\atop g_1+g_2=g, n_1+n_2=n+2}
{\bf S}\left[ \{\RR_{g_1,n_1}, \Delta_{g_2,n_2}\}\right]\, .
\ee
Intuitively this means that $\Gamma_{g,n}$, which encodes the difference between
$\RR_{g,n}(t+\delta t)$ and $\RR_{g,n}(t)$ at the boundary, gets contribution from two
sources to order $\delta t$ 
-- one that encodes the difference between $\RR_{g_1,n_1}(t+\delta t)$ and
$\RR_{g_1,n_1}(t)$ and the other that encodes the difference between
$\RR_{g_2,n_2}(t+\delta t)$ and
$\RR_{g_2,n_2}(t)$. Since both give equal 
contributions after summing over $g_1,g_2,n_1,n_2$ we have kept only one of the
terms and multiplied the result by a factor of 2.

We are now ready to discuss the change in the action. This is given by
\ben \label{efirstterm}
\delta S &=& \delta t \sum_{g=0}^\infty  g_s{}^{2g-2}  \sum_{n=1}^\infty {1\over n!} \, 
 \int_{\delta\RR_{g,n}(t)} \Omega^{(g,n)}_{6g-6+2n}(|\Psi\rangle^{\otimes n})
\nonumber \\ &=& \sum_{g=0}^\infty  g_s{}^{2g-2}  \sum_{n=1}^\infty {1\over n!} \, 
\left[\int_{\Delta_{g,n}(t)} d\Omega^{(g,n)}_{6g-6+2n}(|\Psi\rangle^{\otimes n})
- \int_{\Gamma_{g,n}} \Omega^{(g,n)}_{6g-6+2n}(|\Psi\rangle^{\otimes n})
\right]
\, ,
\een
where in arriving that the last expression we have used 
\refb{epdgn}.
This is the equation analogous to \refb{edeltaaction}.
Since this change is infinitesimal we can  
proceed as earlier and show that this change in the action can be reinterpreted
as the result of a field redefinition of the form
$|\Psi\rangle$ to
$|\Psi\rangle+|\tilde\delta \Psi\rangle$ where $|\tilde\delta\Psi\rangle$ is given by
\be \label{edeltafieldnew}
\langle \Phi| c_0^- |\tilde\delta \Psi\rangle 
=- \sum_{g=0}^\infty g_s{}^{2g} \, \sum_{n=1}^\infty  {1\over (n-1)!} \, 
\int_{\Delta_{g,n}(t)} \, \Omega^{(g,n)}_{6g-5+2n}(|\Phi\rangle, 
|\Psi\rangle^{\otimes (n-1)})\, ,
\ee
for any state $|\Phi\rangle$ in $\HH_0$. The proof of this follows in a 
straightforward fashion, with the change in the term proportional to
$\langle \Psi|c_0^- Q_B |\Psi\rangle$ in the action 
matching the first term on the right hand
side of \refb{efirstterm} and the change in the rest of the terms in the
action matching the
second term on the right hand side of \refb{efirstterm}.

The important aspect of \refb{edeltafieldnew} is that even though the
subspace $\Delta_{g,n}$ contains spurious singularities, the rules of vertical
integration given in \cite{1408.0571} give a 
procedure for integrating $\Omega^{(g,n)}_{6g-5+2n}$
over $\Delta_{g,n}$ yielding a finite result. Thus the field redefinition described in
\refb{edeltafieldnew} is finite.

%\small

%\baselineskip 14pt


\begin{thebibliography}{99}

\bibitem{wittenssft} 
  E.~Witten,
  ``Interacting Field Theory of Open Superstrings,''
  Nucl.\ Phys.\ B {\bf 276}, 291 (1986).
  %%CITATION = NUPHA,B276,291;%%
  %449 citations counted in INSPIRE as of 08 Jul 2014

\bibitem{9202087} 
  R.~Saroja and A.~Sen,
  ``Picture changing operators in closed fermionic string field theory,''
  Phys.\ Lett.\ B {\bf 286}, 256 (1992)
  [hep-th/9202087].
  %%CITATION = HEP-TH/9202087;%%
  %14 citations counted in INSPIRE as of 24 May 2014

\bibitem{9503099}
  N.~Berkovits,
  ``SuperPoincare invariant superstring field theory,''
  Nucl.\ Phys.\ B {\bf 450} (1995) 90
   [Erratum-ibid.\ B {\bf 459} (1996) 439]
  [hep-th/9503099].
  %%CITATION = HEP-TH/9503099;%%

\bibitem{0109100}
  N.~Berkovits,
  ``The Ramond sector of open superstring field theory,''
  JHEP {\bf 0111} (2001) 047
  [hep-th/0109100].
  %%CITATION = HEP-TH/0109100;%%
  %45 citations counted in INSPIRE as of 08 Jul 2014

\bibitem{0406212}
  Y.~Okawa and B.~Zwiebach,
  ``Heterotic string field theory,''
  JHEP {\bf 0407} (2004) 042
  [hep-th/0406212].
  %%CITATION = HEP-TH/0406212;%%
  %21 citations counted in INSPIRE as of 08 Jul 2014

\bibitem{0409018}
  N.~Berkovits, Y.~Okawa and B.~Zwiebach,
  ``WZW-like action for heterotic string field theory,''
  JHEP {\bf 0411} (2004) 038
  [hep-th/0409018].
  %%CITATION = HEP-TH/0409018;%%
  %27 citations counted in INSPIRE as of 08 Jul 2014

\bibitem{0911.2962} 
  M.~Kroyter,
  ``Superstring field theory in the democratic picture,''
  Adv.\ Theor.\ Math.\ Phys.\  {\bf 15}, 741 (2011)
  [arXiv:0911.2962 [hep-th]].
  %%CITATION = ARXIV:0911.2962;%%

\bibitem{1201.1761}
  M.~Kroyter, Y.~Okawa, M.~Schnabl, S.~Torii and B.~Zwiebach,
  ``Open superstring field theory I: gauge fixing, ghost structure, and propagator,''
  JHEP {\bf 1203} (2012) 030
  [arXiv:1201.1761 [hep-th]].
  %%CITATION = ARXIV:1201.1761;%%
  %14 citations counted in INSPIRE as of 08 Jul 2014

\bibitem{1303.2323}
  B.~Jurco and K.~Muenster,
  ``Type II Superstring Field Theory: Geometric Approach and Operadic Description,''
  JHEP {\bf 1304} (2013) 126
  [arXiv:1303.2323 [hep-th]].
  %%CITATION = ARXIV:1303.2323;%%
  %7 citations counted in INSPIRE as of 08 Jul 2014

\bibitem{1312.1677}
  Y.~Iimori, T.~Noumi, Y.~Okawa and S.~Torii,
  ``From the Berkovits formulation to the Witten formulation in open superstring field theory,''
  JHEP {\bf 1403} (2014) 044
  [arXiv:1312.1677 [hep-th]].
  %%CITATION = ARXIV:1312.1677;%%
  %2 citations counted in INSPIRE as of 08 Jul 2014

\bibitem{1312.2948}
  T.~Erler, S.~Konopka and I.~Sachs,
  ``Resolving Witten`s superstring field theory,''
  JHEP {\bf 1404} (2014) 150
  [arXiv:1312.2948 [hep-th]];
  %%CITATION = ARXIV:1312.2948;%%
  %3 citations counted in INSPIRE as of 08 Jul 2014
``NS-NS Sector of Closed Superstring Field Theory,''
  arXiv:1403.0940 [hep-th].
  %%CITATION = ARXIV:1403.0940;%%


\bibitem{1312.7197}
  H.~Kunitomo,
  ``The Ramond Sector of Heterotic String Field Theory,''
  PTEP {\bf 2014} 4,  043B01
  [arXiv:1312.7197 [hep-th]].
  %%CITATION = ARXIV:1312.7197;%%
  %3 citations counted in INSPIRE as of 08 Jul 2014

\bibitem{1403.0940} 
  T.~Erler, S.~Konopka and I.~Sachs,
  ``NS-NS Sector of Closed Superstring Field Theory,''
  arXiv:1403.0940 [hep-th].
  %%CITATION = ARXIV:1403.0940;%%
  %1 citations counted in INSPIRE as of 09 Jun 2014


\bibitem{1311.1257} 
  R.~Pius, A.~Rudra and A.~Sen,
  ``Mass Renormalization in String Theory: Special States,''
  arXiv:1311.1257 [hep-th].
  %%CITATION = ARXIV:1311.1257;%%

\bibitem{1401.7014} 
  R.~Pius, A.~Rudra and A.~Sen,
  ``Mass Renormalization in String Theory: General States,''
  arXiv:1401.7014 [hep-th].
  %%CITATION = ARXIV:1401.7014;%%
  %1 citations counted in INSPIRE as of 12 Mar 2014
  
\bibitem{1404.6254} 
  R.~Pius, A.~Rudra and A.~Sen,
  ``String Perturbation Theory Around Dynamically Shifted Vacuum,''
  arXiv:1404.6254 [hep-th].
  %%CITATION = ARXIV:1404.6254;%%

\bibitem{1408.0571} 
  A.~Sen,
  ``Off-shell Amplitudes in Superstring Theory,''
  arXiv:1408.0571v3 [hep-th].
  %%CITATION = ARXIV:1408.0571;%%

\bibitem{FMS} 
  D.~Friedan, E.~J.~Martinec and S.~H.~Shenker,
  ``Conformal Invariance, Supersymmetry and String Theory,''
  Nucl.\ Phys.\ B {\bf 271}, 93 (1986).
  %%CITATION = NUPHA,B271,93;%%

\bibitem{Verlinde:1987sd} 
  E.~P.~Verlinde and H.~L.~Verlinde,
  ``Multiloop Calculations in Covariant Superstring Theory,''
  Phys.\ Lett.\ B {\bf 192}, 95 (1987).
  %%CITATION = PHLTA,B192,95;%%

\bibitem{lechtenfeld}
O.~Lechtenfeld,
``Superconformal Ghost Correlations On Riemann Surfaces,''
  Phys.\ Lett.\ B {\bf 232}, 193 (1989).
  %%CITATION = PHLTA,B232,193;%%


\bibitem{morozov}
  A.~Morozov,
  ``STRAIGHTFORWARD PROOF OF LECHTENFELD'S FORMULA FOR BETA, gamma CORRELATOR,''
  Phys.\ Lett.\ B {\bf 234}, 15 (1990)
  [Yad.\ Fiz.\  {\bf 51}, 301 (1990)]
  [Sov.\ J.\ Nucl.\ Phys.\  {\bf 51}, 190 (1990)].
  %%CITATION = PHLTA,B234,15;%%

\bibitem{Belopolsky} 
  A.~Belopolsky,
  ``De Rham cohomology of the supermanifolds and superstring BRST cohomology,''
  Phys.\ Lett.\ B {\bf 403}, 47 (1997)
  [hep-th/9609220];
  %%CITATION = HEP-TH/9609220;%%
  %12 citations counted in INSPIRE as of 31 Oct 2013
``New geometrical approach to superstrings,''
  hep-th/9703183;
  %%CITATION = HEP-TH/9703183;%%
  %15 citations counted in INSPIRE as of 31 Oct 2013

\bibitem{9706033} 
  A.~Belopolsky,
``Picture changing operators in supergeometry and superstring theory,''
  hep-th/9706033.
  %%CITATION = HEP-TH/9706033;%%
  %12 citations counted in INSPIRE as of 31 Oct 2013

\bibitem{dp}
  E.~D'Hoker and D.~H.~Phong,
  ``Two loop superstrings. I. Main formulas,''
  Phys.\ Lett.\ B {\bf 529}, 241 (2002)
  [hep-th/0110247].
  %%CITATION = HEP-TH/0110247;%%
``II. The Chiral measure on moduli space,''
  Nucl.\ Phys.\ B {\bf 636}, 3 (2002)
  [hep-th/0110283].
  %%CITATION = HEP-TH/0110283;%%
``III. Slice independence and absence of ambiguities,''
  Nucl.\ Phys.\ B {\bf 636}, 61 (2002)
  [hep-th/0111016].
  %%CITATION = HEP-TH/0111016;%%
``IV: The Cosmological constant and modular forms,''
  Nucl.\ Phys.\ B {\bf 639}, 129 (2002)
  [hep-th/0111040].
  %%CITATION = HEP-TH/0111040;%%
``V. Gauge slice independence of the N-point function,''
  Nucl.\ Phys.\ B {\bf 715}, 91 (2005)
  [hep-th/0501196].
  %%CITATION = HEP-TH/0501196;%%
``VI: Non-renormalization theorems and the 4-point function,''
  Nucl.\ Phys.\ B {\bf 715}, 3 (2005)
  [hep-th/0501197].
  %%CITATION = HEP-TH/0501197;%%
``VII. Cohomology of Chiral Amplitudes,''
  Nucl.\ Phys.\ B {\bf 804}, 421 (2008)
  [arXiv:0711.4314 [hep-th]].
  %%CITATION = ARXIV:0711.4314;%%


\bibitem{1209.5461} 
  E.~Witten,
  ``Superstring Perturbation Theory Revisited,''
  arXiv:1209.5461 [hep-th].
  %%CITATION = ARXIV:1209.5461;%%
  %6 citations counted in INSPIRE as of 27 Feb 2013


\bibitem{1304.2832}
E.~Witten,
``More On Superstring Perturbation Theory,''
  arXiv:1304.2832 [hep-th].
  %%CITATION = ARXIV:1304.2832;%%
  %9 citations counted in INSPIRE as of 31 Oct 2013


\bibitem{Witten}
E.~Witten,
  ``Notes On Supermanifolds and Integration,''
  arXiv:1209.2199 [hep-th];
  %%CITATION = ARXIV:1209.2199;%%
  %15 citations counted in INSPIRE as of 31 Oct 2013
 ``Notes On Super Riemann Surfaces And Their Moduli,''
  arXiv:1209.2459 [hep-th];
  %%CITATION = ARXIV:1209.2459;%%
  %14 citations counted in INSPIRE as of 31 Oct 2013
``Notes On Holomorphic String And Superstring Theory Measures Of Low Genus,''
  arXiv:1306.3621 [hep-th];
  %%CITATION = ARXIV:1306.3621;%%
  %2 citations counted in INSPIRE as of 31 Oct 2013

\bibitem{donagi-witten} 
  R.~Donagi and E.~Witten,
  ``Supermoduli Space Is Not Projected,''
  arXiv:1304.7798 [hep-th];
``Super Atiyah classes and obstructions to splitting of supermoduli space,''
  arXiv:1404.6257 [hep-th].
  %%CITATION = ARXIV:1404.6257;%%
  %%CITATION = ARXIV:1304.7798;%%
  %8 citations counted in INSPIRE as of 28 Oct 2013

  \bibitem{1403.5494} 
  E.~D'Hoker and D.~H.~Phong,
  ``Two-loop vacuum energy for Calabi-Yau orbifold models,''
  Nucl.\ Phys.\ B {\bf 877}, 343 (2013)
  [arXiv:1307.1749];
  %%CITATION = ARXIV:1307.1749;%%
  %2 citations counted in INSPIRE as of 25 Mar 2014
  E.~D'Hoker 
  Topics in Two-Loop Superstring Perturbation Theory
arXiv:1403.5494 [hep-th];
  %%CITATION = ARXIV:1403.5494;%%

\bibitem{catoptric} 
  J.~J.~Atick, G.~W.~Moore and A.~Sen,
  ``Catoptric Tadpoles,''
  Nucl.\ Phys.\ B {\bf 307}, 221 (1988).
  %%CITATION = NUPHA,B307,221;%%
  %71 citations counted in INSPIRE as of 24 Mar 2014

\bibitem{wittensft} 
  E.~Witten,
  ``Noncommutative Geometry and String Field Theory,''
  Nucl.\ Phys.\ B {\bf 268}, 253 (1986).
  %%CITATION = NUPHA,B268,253;%%
  
\bibitem{saadi} 
  M.~Saadi and B.~Zwiebach,
  ``Closed String Field Theory from Polyhedra,''
  Annals Phys.\  {\bf 192}, 213 (1989).
  %%CITATION = APNYA,192,213;%%
  %146 citations counted in INSPIRE as of 06 Jul 2014

 \bibitem{kugo}
T.~Kugo and K.~Suehiro,
``Nonpolynomial Closed String Field Theory,''
  Phys.\ Lett.\ B {\bf 226}, 48 (1989);
  %%CITATION = PHLTA,B226,48;%%
``Nonpolynomial Closed String Field Theory: Action and Its Gauge Invariance,''
  Nucl.\ Phys.\ B {\bf 337}, 434 (1990).
  %%CITATION = NUPHA,B337,434;%%


\bibitem{9206084} 
  B.~Zwiebach,
  ``Closed string field theory: Quantum action and the B-V master equation,''
  Nucl.\ Phys.\ B {\bf 390}, 33 (1993)
  [hep-th/9206084].
  %%CITATION = HEP-TH/9206084;%%
  %335 citations counted in INSPIRE as of 24 May 2014

\bibitem{aseq} 
  A.~Sen,
  ``Equations of Motion in Nonpolynomial Closed String Field Theory and Conformal Invariance of Two-dimensional Field Theories,''
  Phys.\ Lett.\ B {\bf 241}, 350 (1990).
  %%CITATION = PHLTA,B241,350;%%

\bibitem{9301097} 
  H.~Hata and B.~Zwiebach,
  ``Developing the covariant Batalin-Vilkovisky approach to string theory,''
  Annals Phys.\  {\bf 229}, 177 (1994)
  [hep-th/9301097].
  %%CITATION = HEP-TH/9301097;%%

\bibitem{nelson} 
  P.~C.~Nelson,
  ``Covariant Insertion of General Vertex Operators,''
  Phys.\ Rev.\ Lett.\  {\bf 62}, 993 (1989).
  %%CITATION = PRLTA,62,993;%%
  %39 citations counted in INSPIRE as of 31 Oct 2013
  
\bibitem{unpub}
A.~Sen and E.~Witten, unpublished.

\bibitem{berera} 
  A.~Berera,
  ``Unitary string amplitudes,''
  Nucl.\ Phys.\ B {\bf 411}, 157 (1994).
  %%CITATION = NUPHA,B411,157;%%

\bibitem{1307.5124}
E.~Witten,
``The Feynman $i \epsilon$ in String Theory,''
  arXiv:1307.5124 [hep-th].
  %%CITATION = ARXIV:1307.5124;%%
  %1 citations counted in INSPIRE as of 31 Oct 2013

\bibitem{mukherji} 
  S.~Mukherji and A.~Sen,
  ``Some all order classical solutions in nonpolynomial closed string field theory,''
  Nucl.\ Phys.\ B {\bf 363}, 639 (1991).
  %%CITATION = NUPHA,B363,639;%%

\end{thebibliography}
\end{document}